\documentclass[acmtog,authorversion]{acmart}

\usepackage{bm}
\usepackage{bbm}

\usepackage{graphicx}
\usepackage{tabularx}
\usepackage{makecell}
\usepackage{subcaption}
\usepackage{multirow}
\usepackage{amsmath}
\usepackage{mathrsfs}

\newcommand{\vect}[1]{\bm{#1}}

\newcommand{\hatvect}[1]{\vect{\hat{#1}}}
\newcommand{\overlinevect}[1]{\vect{\overline{#1}}}

\newcommand{\eqword}[1]{{\text{#1}}}

\newcommand{\gesLabel}[1]{{\texttt{#1}}}

\usepackage{xspace}

\newcommand{\keep}[1]{}
\newcommand{\old}[1]{}

\DeclareMathOperator*{\argmin}{arg\,min}

\DeclareMathOperator*{\stopgrad}{sg}

\usepackage{url}
\usepackage{hyperref}
\newcommand{\fig}{Figure{}~}

\AtBeginDocument{%
  \providecommand\BibTeX{{%
    \normalfont B\kern-0.5em{\scshape i\kern-0.25em b}\kern-0.8em\TeX}}}

\setcopyright{acmlicensed}
\acmJournal{TOG}
\acmYear{2024} 
\acmVolume{43} 
\acmNumber{4} 
\acmArticle{}
\acmMonth{7}
\acmDOI{10.1145/3658134}

\acmSubmissionID{0}

\begin{document}

\title{Semantic Gesticulator: Semantics-Aware Co-Speech Gesture Synthesis}

\author{Zeyi Zhang}
\authornote{these authors contributed equally to this research.}
\email{illusence1@gmail.com}
\affiliation{%
    \institution{School of Electronics Engineering and Computer Science, Peking University}
    \country{China}
}

\author{Tenglong Ao}
\authornotemark[1]
\email{aubrey.tenglong.ao@gmail.com}
\affiliation{%
    \institution{School of Computer Science, Peking University}
    \country{China}
}

\author{Yuyao Zhang}
\authornotemark[1]
\email{2020201710@ruc.edu.cn}
\affiliation{%
    \institution{Renmin University of China}
    \country{China}
}

\author{Qingzhe Gao}
\email{gaoqingzhe97@gmail.com}
\affiliation{%
    \institution{Shandong University}
    \country{China}
}
\affiliation{%
    \institution{Peking University}
    \country{China}
}

\author{Chuan Lin}
\email{chuanlin2015@foxmail.com}
\affiliation{%
    \institution{Peking University}
    \country{China}
}

\author{Baoquan Chen}
\email{baoquan@pku.edu.cn}
\orcid{0000-0003-4702-036X} 
\affiliation{%
  \institution{Peking University}
  \country{China}
}
\affiliation{%
  \institution{State Key Lab of General AI}
  \country{China}
}

\author{Libin Liu}
\authornote{corresponding author}
\email{libin.liu@pku.edu.cn}
\orcid{0000-0003-2280-6817}
\affiliation{%
  \institution{Peking University}
  \country{China}
}
\affiliation{%
  \institution{State Key Lab of General AI}
  \country{China}
}


\begin{abstract}
  
    In this work, we present \emph{Semantic Gesticulator}, a novel framework designed to synthesize realistic gestures accompanying speech with strong semantic correspondence. Semantically meaningful gestures are crucial for effective non-verbal communication, but such gestures often fall within the long tail of the distribution of natural human motion. The sparsity of these movements makes it challenging for deep learning-based systems, trained on moderately sized datasets, to capture the relationship between the movements and the corresponding speech semantics. To address this challenge, we develop a generative retrieval framework based on a large language model. This framework efficiently retrieves suitable semantic gesture candidates from a motion library in response to the input speech. To construct this motion library, we summarize a comprehensive list of commonly used semantic gestures based on findings in linguistics, and we collect a high-quality motion dataset encompassing both body and hand movements. We also design a novel GPT-based model with strong generalization capabilities to audio, capable of generating high-quality gestures that match the rhythm of speech. Furthermore, we propose a semantic alignment mechanism to efficiently align the retrieved semantic gestures with the GPT's output, ensuring the naturalness of the final animation. Our system demonstrates robustness in generating gestures that are rhythmically coherent and semantically explicit, as evidenced by a comprehensive collection of examples. User studies confirm the quality and human-likeness of our results, and show that our system outperforms state-of-the-art systems in terms of semantic appropriateness by a clear margin. We will release the code and dataset for academic research.
        
\end{abstract}



\begin{teaserfigure}
  \centering
  \includegraphics[width=\textwidth]{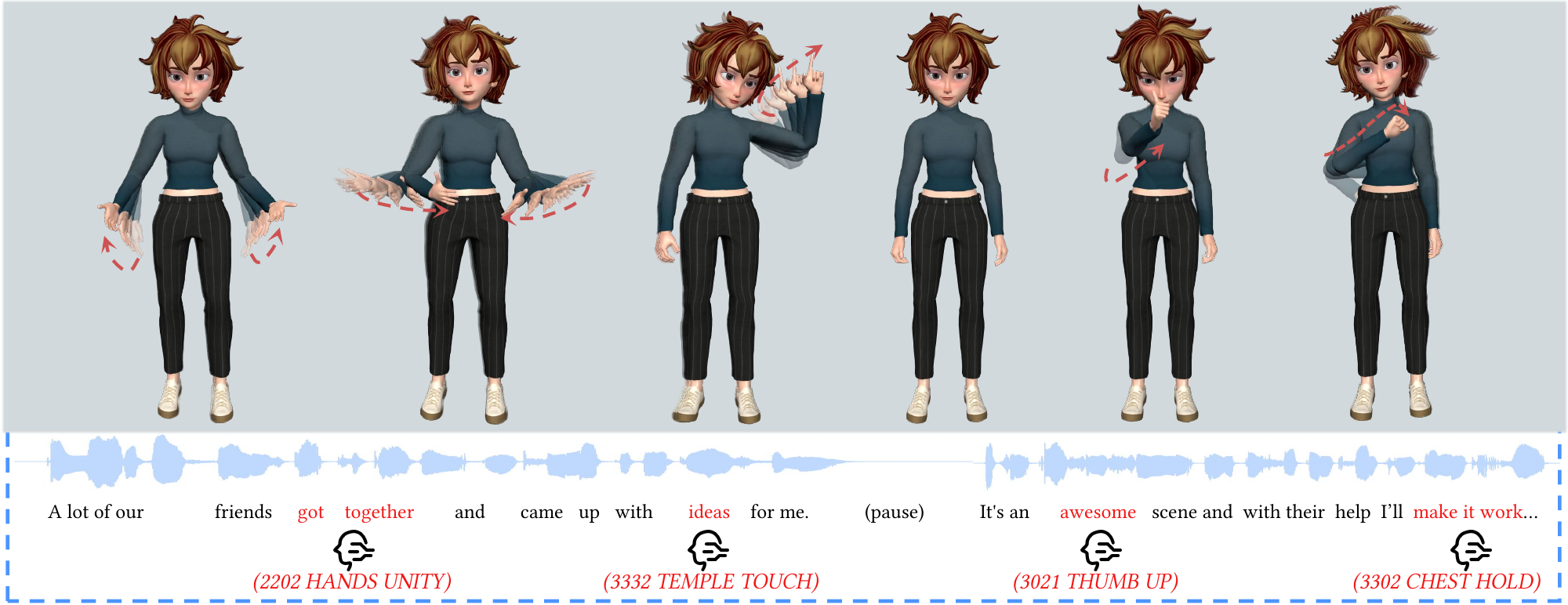}
  \caption{Our system can synthesize realistic co-speech gestures with strong rhythmic coherence and semantic correspondence.}
  \Description{}
  \label{fig:teaser}
\end{teaserfigure}

\maketitle

\section{Introduction}
\label{sec:introduction}
Gestures, which are hand and body movements that are non-verbal and non-manipulative, often accompany speech and play an important role in communication~\cite{mcneill1992hand}. They not only enhance expression, making it more dynamic, especially through rhythmic movements known as \emph{beat gestures}, but also increase communicative efficacy. This includes pointing or deictic gestures that convey specific references, metaphoric and iconic gestures that depict concepts, and emblems that symbolize specific words or phrases~\cite{nyatsanga2023data_driven_gesture_survey}. In this work, we refer to gestures with such communicative functions as \emph{semantic gestures}. The occurrence of such gestures is much less frequent than that of beat gestures and falls within the long tail distribution of gestures.~\cite{liu2021beatdataset,liu2022disco}.

Deep learning-based approaches have recently achieved significant success in synthesizing natural-looking gestures from speech input. These methods often excel at producing beat gestures that are in rhythmic harmony with the speech, but tend to stuggle with semantic gestures that reflect the meaning of the spoken words. Many previous deep learning-based systems adopt textual features extracted from transcripts as inputs to perceive semantics~\cite{kucherenko2020gesticulator,yoon2020trimodalgesture,ao2022rhythmicgesticulator,pang2023bodyformer,zhi2023livelyspeaker}. However, since the majority of words in transcripts are not accompanied by semantically meaningful gestures, the learning process often tends to overlook these sparse semantic gestures and treat them as noises in the training data.
There are methods attempting to refine the training procedure to address such issues. For example, \citet{liu2022disco} cluster gestures and resample within each cluster to improve data balance. \citet{ao2023gesturediffuclip} attempt to discover implicit correspondence between semantic gestures and semantically salient words using contrastive learning. Meanwhile, \citet{liang2022seeg} train classifiers to classify gestures into a small number of labels and use these classifiers as extra perceptual objectives. Although these methods show effectiveness in certain scenarios, they still rely on sufficient coverage of semantic gesture-speech pairs within training data. Given the sparsity of such data, obtaining this coverage is challenging. Consequently, these methods struggle in practice to reliably produce appropriate semantic gestures at the correct moments, especially when dealing with a diverse range of semantic gesture categories.

A widely-used and robust strategy for enabling neural systems to perceive sparse and hard-to-mine data clues is retrieval augmentation, which has achieved great success in natural language processing~\cite{lewis2020rag,izacard2022ragfewshot,ram2023ragincontext}. The core idea involves retrieving useful items from an external database to enhance generation. When such retrieval is conducted by a powerful model, like a large language model, the external database effectively becomes augmented in return by the knowledge embedded in the retrieval model. This knowledge, originating from a much larger volume of data and potentially from different modalities, extends the limits of the data being retrieved.

Adapting this idea to semantic gestures is non-trivial, containing two challenges: 1) establishing a high-quality motion dataset with sufficient diversity to cover commonly used semantic gestures; 2) developing a retrieval model adept at choosing appropriate semantic gestures according to context and determining their timing. To address the first challenge, we compile a comprehensive set of semantic gestures commonly used in human communication, drawing on relevant linguistic and human behavioral studies (e.g. \cite{Morris1994BodytalkAW,Wagner2003FieldGT,Kipp2005GestureGB,GESTUNO}). Based on this collection, we record a high-quality dataset of body, hand, and finger movement using motion capture. This dataset contains over 200 types of semantic gestures, each offering a variety of gestures. As for the second challenge, large language models like ChatGPT~\cite{openai2022chatgpt} possess strong contextual understanding and generalization capabilities. We construct a generative retrieval model by fine-tuning a large language model, enabling it to efficiently retrieve appropriate semantic gestures based on input speech in an end-to-end manner and determine their timing.

We refer to this framework as \emph{Semantic Gesticulator}, a co-speech gesture synthesis neural system designed to generate diverse and appropriate semantic gestures while ensuring rhythmic coherence with speech. We learn a GPT-based~\cite{radford2018gpt1} gesture generative model to accommodate the retrieved gestures. This model is built upon discrete tokens extracted by a scalable, body part-aware Residual VQ-VAE~\cite{zeghidour2022soundstream}. It generalizes to a broad range of audio inputs, resulting in gestures that are both realistic and rhythmically in sync with the input speech. Moreover, we develop a semantics-aware gesture alignment mechanism. This mechanism fuses semantic and rhythmic gestures at the latent space level, ensuring that the generated gestures are not only meaningful but also exhibit rhythmic harmony.

In summary, the technical contributions of this work include:
\begin{itemize}
    \item We introduce a novel semantics-aware co-speech gesture synthesis system that produces natural and semantically rich gestures. The GPT-based generator and the semantics-aware alignment mechanism effectively ensure motion quality and generalization across different audio inputs.
    \item We develop an LLM-based generative semantic gesture retrieval framework capable of efficiently retrieving semantic gestures from a gesture library.
    \item We compile a comprehensive list of commonly used semantic gestures and capture a high-quality dataset according to it. The list and the dataset will be released to the community for academic research.
\end{itemize}
\section{Related Work}
\label{sec:related_work}
\subsection{Co-Speech Gesture Synthesis}
Early methods for co-speech gesture synthesis is rule-based approaches, which primarily employ linguistic mapping rules to convert speech into sequences of pre-defined gesture clips~\cite{cassell1994rulefullbody,cassell2001beat,kipp2004gesture,kopp2006bml}. A thorough review of these methods is presented by \cite{wagner2014rulereview}. Although rule-based methods yield interpretable and controllable outcomes, the construction of mapping rules and gesture databases is labor-intensive and movements generated by these methods are not realistic. To reduce the manual effort involved in creating rules and improve motion quality, data-driven approaches have increasingly become the dominant methodology in this field. A comprehensive survey of these methods is provided by \cite{nyatsanga2023data_driven_gesture_survey}. Statistical models are initially employed to extract mapping rules from data~\cite{neff2008videogesture,levine2009prosodygesture,levine2010gesturecontroller}, but they still require a meticulously curated library of gesture units to generate the final motion. Subsequently, the advanced capabilities of deep neural networks enable the training of complex end-to-end models directly using raw gesture data. Deep learning-based systems can be categorized based on the randomness of their output. One choice is deterministic models, which are mainly based on MLPs~\cite{kucherenko2020gesticulator}, CNNs~\cite{habibie2021videogesture}, RNNs~\cite{yoon2019robot,yoon2020trimodalgesture,bhattacharya2021affectivegesture,liu2022hierarchicalgesture,ghorbani2022zeroeggs,liu2022disco,liu2022disco,xu2023chainofgeneration,liang2022seeg}, and Transformer~\cite{bhattacharya2021text2gestures,voss2023augmented,qi2023emotiongesturenetease,qi2023weaklyemotiongesture}. Gestures generated by deterministic methods easily converge to ``mean" poses due to the inherent many-to-many relationship between speech and gestures. Generative models are widely used to alleviate this problem, such as VAEs~\cite{li2021audio2gesture,ghorbani2022zeroeggs}, VQ-VAEs~\cite{yi2022talkshow,yazdian2022gesture2vec,liu2022vqgesturevideo,zhou2023unifiedxiaobing,zhang2023speechact,liu2023emage,ao2022rhythmicgesticulator,ng2024audio2photoreal}, flow-based models~\cite{alexanderson2020stylegesture,ye2022styleflowgesture,kucherenko2021speech2properties2gestures}, and diffusion-based models~\cite{alexanderson2023listendenoiseaction,zhang2023diffmotion,zhang2023audioallinone,xue2023conversationalgesture,ji2023c2g2,ao2023gesturediffuclip,mehta2023diffttsg,yang2023diffusestylegesture,deichler2023diffusionjoint,yin2023emog,chhatre2023emotionalgesture,zhi2023livelyspeaker,zhu2023tamingdiffusiongesture,chen2023diffsheg,yang2024freetalker}. Additionally, some hybrid systems integrate deep features with motion graphs~\cite{zhou2022gesturemaster} or motion matching~\cite{yang2023qpgesture,habibie2022motionmatching,ExpressGesture} to enhance the controllability of the system. Generating meaningful gestures in sync with speech robustly is challenging for neural systems~\cite{yoon2022genea}. \citet{liang2022seeg} design specific semantic gesture classifiers to explicitly guide the generator. Some systems opt to mine implicit content as a representation of semantics~\cite{liu2022disco,ao2023gesturediffuclip}. 
\citet{gao2023gesgpt} define six general semantic gestures and employ a Large Language Model (LLM) as a simple classifier. Through prompt engineering, it identifies corresponding gestures for each sentence of text. Finally, the identified gestures are merged with gestures generated by the neural system through linear interpolation. There are there key differences between this work with our system: (a) our method enables generating over 200 types of semantic gestures, which covers commonly used scenarios; (b) we develop a generative retrieval framework through fine-tuning the LLM, capable of efficiently retrieving semantic gestures from a large gesture library; and (c) a semantics-aware gesture alignment mechanism is proposed to fuse semantic and rhythmic gestures at the latent space
level, ensuring that the generated gestures are both meaningful and rhythm-coherent.

\subsection{Speech-Gesture Dataset}
Current co-speech gesture datasets could be divided into two types: pose-estimated and motion-captured. For the former, \citet{ginosar2019stylegesture} propose the Speech2Gesture Dataset, which employs OpenPose~\cite{cao2017openpose} to extract 2D poses from News and Teaching videos. This dataset is then lifted to 3D pose~\cite{habibie2021videogesture} and SMPL-X~\cite{yi2022talkshow}. Similarly, \citet{yoon2019robot} estimate 2D poses from TED videos and build the TED Dataset, which is also extended to 3D pose~\cite{yoon2020trimodalgesture,liu2022hierarchicalgesture} and SMPL-X~\cite{lu2023tedsmplx}. Although pose-estimation methods facilitate the extraction of vast amounts of data from videos, their accuracy remains constrained. Motion-captured methods can yield high-quality motion datasets but are often costly. The Trinity Dataset~\cite{ferstl2018trinity} showcases a male actor with $4$ hours of data, and the TalkingWithHands Dataset~\cite{lee2019talkingwithhands} gathers data from conversational scenarios involving two speakers. The BEAT Dataset~\cite{liu2021beatdataset} first incorporates both 3D pose and facial blendshapes. \citet{liu2023emage} add mesh-level data to this dataset. ZEGGS~\cite{ghorbani2022zeroeggs} focuses on a single speaker across 12 different styles. In this work, we collect a motion-captured gesture dataset encompassing commonly used semantic gestures, designed to augment and enrich existing datasets.

\subsection{Generative Retrieval}
Unlike the traditional index-retrieval-rank paradigm~\cite{LLM4IR}, generative retrieval involves storing knowledge in model parameters~\cite{LLM4IR,DSI}. It focuses on generating index identifiers, such as numbers and titles, sequentially through an autoregressive fashion to achieve end-to-end retrieval~\cite{GENRE}. This model-based information retrieval has recently gained significant attention in academia~\cite{Dynamicretriever,CorpusBrain}.
Methods like DSI~\cite{DSI} and NCI~\cite{NCI} encode document content into vectors and use hierarchical clustering to generate numeric, semantically-based identifiers for retrieval. These number-based identifiers implicitly capture the document's hierarchical information and demonstrate efficient and effective performance. However, converting text to number-based identifiers can result in a loss of semantic detail, posing challenges for neural networks in learning this mapping function.
To transfer the knowledge compressed in Pre-trained Language Models (PLMs), GENRE~\cite{GENRE} focuses on retrieving entities by finetuning a T5 model~\cite{T5} to generate term-based identifiers such as each entity's unique title, progressing from left to right and token by token.
Additionally, there are studies exploring the integration of multiple identifiers and how to use Large Language Models (LLMs)~\cite{Multiview,SEAL,ziems2023large}.
We can interpret our semantic gesture synthesis as a specific retrieval task. It uses speech text as a query to retrieve and decode appropriate semantic gesture identifiers at relevant locations using an autoregressive fashion, fulfilling the overall system's requirements. In contrast to sequence labeling tasks, this approach enables cross-encoding of context and considers semantics of gestures.

\section{System Overview}
\label{sec:system_overview}

\begin{figure}[t]
    \centering
    \includegraphics[width=\linewidth]{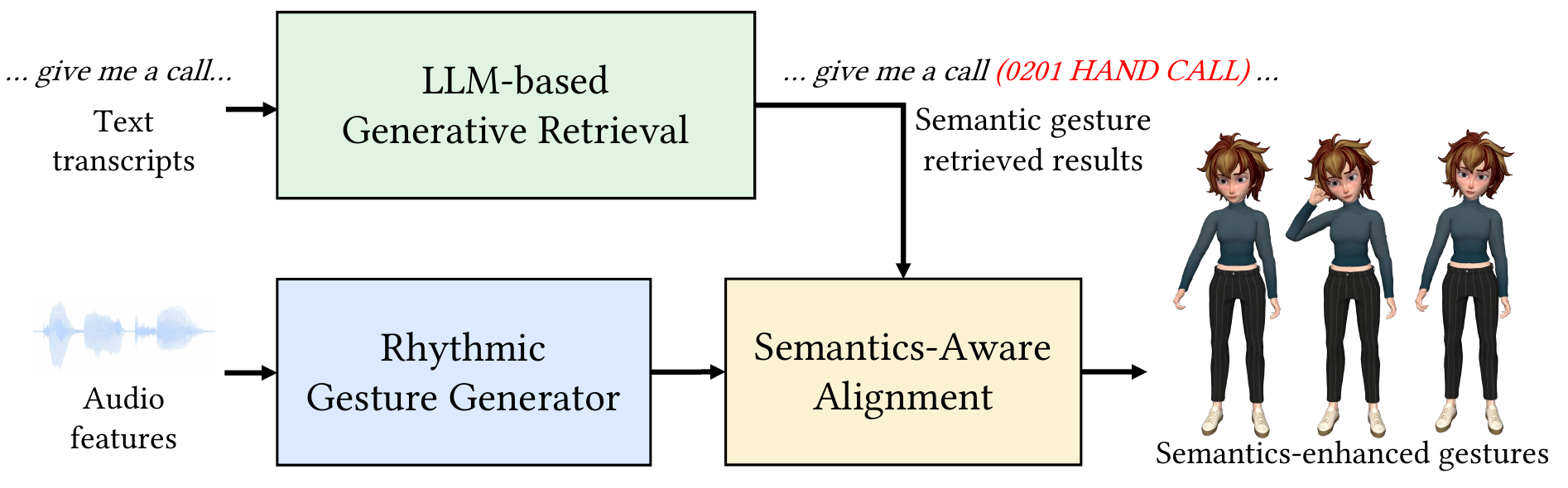}
    \caption{Our system is composed of three principal components: (a) an end-to-end neural generator, adept at handling a wide array of speech audio inputs to create gesture animations that are in rhythm with the speech; (b) a generative retrieval framework based on a large language model (LLM), adept at interpreting transcript context and selecting suitable semantic gestures from an extensive library covering commonly used gestures; and (c) a semantics-aware alignment mechanism, which amalgamates the chosen semantic gestures with the rhythmically produced motion, culminating in gestures that are semantically enriched.}
    \Description{}
    \label{fig:system_overview}
\end{figure}

Our system processes audio and speech transcripts as inputs to generate realistic full-body gestures, including finger motion, that are both rhythmically and semantically aligned with the speech content. It is capable of robustly synthesizing sparse semantic gestures, vital for effective communication.

Our system is built upon a discrete latent motion space, learned through the use of a residual VQ-VAE \cite{zeghidour2022soundstream}. This approach tokenize a sequence of gestures into hierarchical and compact motion tokens, ensuring both motion quality and diversity. As illustrated in \fig\ref{fig:system_overview}, our system comprises three key modules: (a) an end-to-end neural generator capable of processing a diverse range of speech audio inputs to produce rhythm-matched gesture animations utilizing the GPT architecture \cite{GPT2}; (b) a large language model (LLM)-based generative retrieval framework that comprehends the context of the transcript input and retrieves appropriate semantic gestures from a high-quality motion library covering commonly used semantic gestures; and (c) a semantics-aware alignment mechanism that integrates the retrieved semantic gestures with the rhythmic motion generated, resulting in semantically-enhanced gesture animation. In subsequent sections, we detail the components of our system and their respective training processes.
\section{Co-Speech Gesture GPT Model}
\label{sec:pretrained_model}

\begin{figure*}[t]
    \centering
    \includegraphics[width=0.9\textwidth]{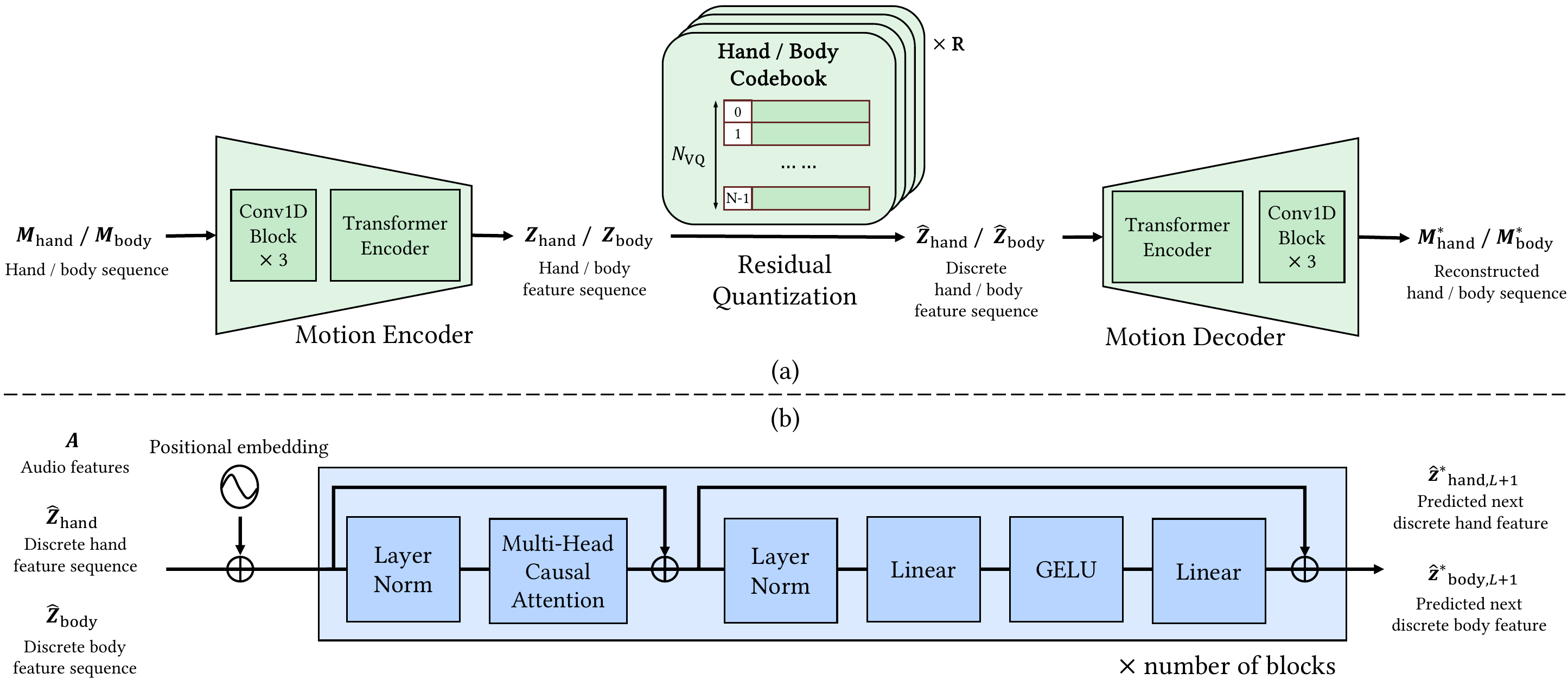}
    \caption{The process of synthesizing a rhythm-coherent gesture segment consists of: (a) a residual VQ-VAE~\cite{zeghidour2022soundstream} learns a hierarchical categorical space to represent motion as discrete tokens; (b) a powerful GPT-based~\cite{GPT2} generator predicts the future gesture token conditioned on the preceding gesture tokens and synchronized audio features in an autoregressive manner.}
    \Description{}
    \label{fig:pre-trained_model}
\end{figure*}

We design a gesture generative model $\mathcal{G}$ enabling synthesizing rhythm-matched gestures as the foundation for the following semantic enhancement. The generator $\mathcal{G}$ predicts a sequence of discrete gesture tokens $\hatvect{Z}=[[\hatvect{z}^r_l]_{r=1}^{R}]_{l=1}^{L}$ conditioned on a speech, where the space of gesture tokens is pre-learned by a Residual VQ-VAE (RVQ)~\cite{zeghidour2022soundstream}, $R$ is the number of the residual quantization layers, $\hatvect{z}^r_l\in\mathbb{R}^{C}$, and $C$ is the dimension of the latent space. Then, the token sequence is decoded into gesture motion $\vect{M}=[\vect{m}_k]_{k=1}^{K}$ using the RVQ decoder $\mathcal{D}_{\eqword{VQ}}$, where the RVQ's downsampling rate is $d=K/L$. Each pose $\vect{m}_k\in\mathbb{R}^{3+6J}$ consists of the translation of the avatar and the rotations of its $J$ joints. The rotations are parameterized as the exponential map. Next, we will discuss the details of the gesture tokenizer and the generator $\mathcal{G}$.

\subsection{Gesture Tokenizer}
\label{subsec:gesture_vq-vae}

\begin{figure}[t]
    \centering
    \includegraphics[width=.9\linewidth]{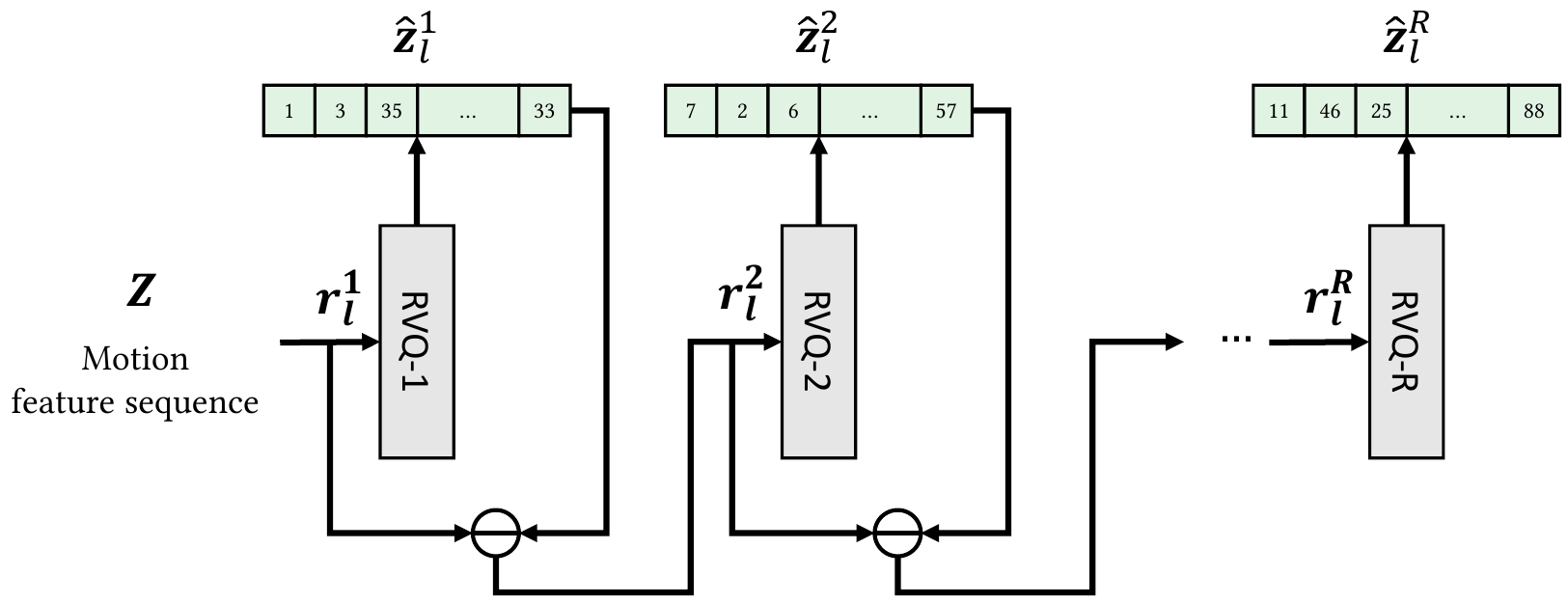}
    \caption{Residual quantization module. The motion features $\vect{Z}$ are iteratively quantized by a series of residual quantization layers. The first RVQ layer is a special RVQ layer, where the preceding residue is $\vect{r}_l^1 = \vect{z}_l$.}
    \Description{}
    \label{fig:rvq}
\end{figure}

Vanilla VQ-VAE~\cite{van2017vqvae} exhibits limited representational capacity, which hinders its ability to reconstruct complex motions, particularly in finger animation. To address this, as shown in \fig\ref{fig:pre-trained_model} (a), we enhance the standard VQ-VAE across three dimensions: 1) by dividing the motion representation into two parts, namely, body and hands, and compressing them independently; 2) by designing a more powerful encoder $\mathcal{E}_{\eqword{VQ}}$ and decoder $\mathcal{D}_{\eqword{VQ}}$ consisting of 1D convolutional layers and the Transformer layer~\cite{vaswani2017transformer}; 3) by improving the capacity of the quantization module through the addition of multiple residual quantization layers.

Specifically, we divide the gesture sequence $\vect{M}$ into the body part $\vect{M}_{\eqword{body}}$ and the hand part $\vect{M}_{\eqword{hand}}$, utilizing two independent RVQ networks to model them respectively. The body part encompasses joints except for the fingers, while the hand part represents finger movements. Each network, acting as an expert, focuses on one specific part, thereby facilitating the handling of the complexity inherent in human movements. For simplicity, we next still use $\vect{M}$ to stand in for the general motion sequence. Formally, the RVQ encoder $\mathcal{E}_{\eqword{VQ}}$ computes the gesture feature sequence $\vect{Z}=[\vect{z}_l]_{l=1}^{L}$ as
\begin{align}
    \vect{Z} = \mathcal{E}_{\eqword{VQ}}(\vect{M}),
\end{align}
where $\vect{z}_l\in\mathbb{R}^{C}$ and $C$ is the dimension of the latent space.

Then, we should quantize $\vect{Z}$ into $\hatvect{Z}$. Due to the high diversity of motion data, the representational capacity of the quantization module needs to be extended. But simply increasing the size of the codebook $\mathcal{C}$ would lead to inefficient and unstable training, e.g., code collapse~\cite{dhariwal2020jukebox}. To address it, inspired by \cite{zeghidour2022soundstream}, we implement a hierarchical architecture with multiple residual vector quantization layers and corresponding independent codebooks $\{\mathcal{C}_i\}_{i=1}^{R}$ to iteratively model the motion features, where the size of each codebook is $N_{\eqword{VQ}}$. As demonstrated in \fig\ref{fig:rvq}, a RVQ layer, e.g., RVQ-$i$, processes the residue $\vect{r}_{l}^{i}$ from the quantization of the preceding RVQ layer and quantizes it into a discrete feature sequence $\hatvect{z}_l^i$ by looking up the corresponding codebook $\mathcal{C}_{i}$ as
\begin{align}
    \hatvect{z}_l^i = \argmin_{\hatvect{z}'\in \mathcal{C}_{i}} \lVert \hatvect{z}' - \vect{r}_{l}^{i} \rVert_2.
\end{align}
When $i < R$, the new residue $\vect{r}_{l}^{i+1}$ is calculated as
\begin{align}
    \vect{r}_{l}^{i+1} = \vect{r}_{l}^{i} - \hatvect{z}_l^i.
\end{align}
This cycle continues sequentially for R times and finally the input motion feature $\vect{z}_l$ is quantized into a hierarchical discrete feature sequence
$[\hatvect{z}_{l}^i]_{i=1}^R$. Notably, the first layer (RVQ-$1$) actually functions as the standard VQ layer. We treat it as a special RVQ layer, where the preceding residue is $\vect{r}_{l}^1 = \vect{z}_l$. As the number of layers increases, the Residual VQ-VAE demonstrates an exponential expansion in its capacity~\cite{yao2023moconvq}. This enhancement significantly boosts the model's capability for expression.

Finally, we compute the reconstructed motion $\vect{M}^*$ as
\begin{align}
    \vect{M}^* = \mathcal{D}_{\eqword{VQ}}(\hatvect{Z}).
\end{align}
Following \cite{van2017vqvae}, the loss function is defined as
\begin{align}
    \mathcal{L}_{\eqword{RVQ}} &= w_{1}\lVert \vect{M}-\vect{M}^{*} \rVert_1 + w_{2}\lVert \vect{\dot M}-\vect{\dot M}^{*} \rVert_1 + w_{2}\lVert \vect{\ddot M}-\vect{\ddot M}^{*} \rVert_1 \nonumber\\
    &+  w_{3}\lVert \mathcal{E}_{\eqword{VQ}}(\vect{M}) -\stopgrad([\sum_{i=1}^{R}{\hatvect{z}_l^i}]_{l=1}^L) \rVert_2^2 \nonumber\\
    &+  w_{4}\lVert \stopgrad(\mathcal{E}_{\eqword{VQ}}(\vect{M})) - {[\sum_{i=1}^{R}{\hatvect{z}_l^i}]_{l=1}^L} \rVert_2^2,
\end{align}
where
$\vect{\dot M}$ and $\vect{\ddot M}$ represent the first-order (velocity) and second-order (acceleration) derivatives of $\vect{M}$ on time. $\stopgrad$ stands for the \emph{stop gradient} operator that prevents the gradient from backpropagating through it. And $[w_i]_{i=1}^{4}$ corresponds to the weighted factors.

\subsection{Gesture Generator}
\label{subsec:gesture_generator}
As shown in \fig\ref{fig:pre-trained_model} (b), the gesture generator $\mathcal{G}$ is based on the GPT-2~\cite{GPT2}, which predicts the future gesture tokens ($\hatvect{z}_{\eqword{hand},L+1}^*$, $\hatvect{z}_{\eqword{body},L+1}^*$)  in an autoregressive manner conditioned on preceding motion tokens 
($\hatvect{Z}_{\eqword{hand}}$ $=$ $[\hatvect{z}_{\eqword{hand},l}]_{l=1}^{L}$, $\hatvect{Z}_{\eqword{body}}$ $=$ $[\hatvect{z}_{\eqword{body},l}]_{l=1}^{L}$) and the synchronized audio features $\vect{A}$ $=$ $[a_l]_{l=1}^{L+1}$. The process is formalized as
\begin{align}
    \hatvect{z}_{\eqword{hand},L+1}^*, \hatvect{z}_{\eqword{body},L+1}^* = \mathcal{G}(\vect{A}, [\hatvect{z}_{\eqword{hand},l}]_{l=1}^{L}, [\hatvect{z}_{\eqword{body},l}]_{l=1}^{L}),
\end{align}
where the audio features $\vect{A}$ consist of \emph{mel frequency cepstral co-efficients (MFCC)}, \emph{MFCC delta}, \emph{constant-Q chromagram}, \emph{onset}, and \emph{tempogram}, which are extracted by the famous audio processing toolbox Librosa~\cite{mcfee2015librosa}. Our approach incorporates the causal attention layer, as introduced by \citet{vaswani2017transformer}, which is designed to facilitate communication exclusively between current and preceding data, thereby maintaining causality. Following \cite{GPT2}, we minimize the standard categorical cross-entropy loss to train the generator.
\section{Generative Semantic Gesture Retrieval}

\begin{figure}[htb]
  \centering
  \includegraphics[width=.7\linewidth]{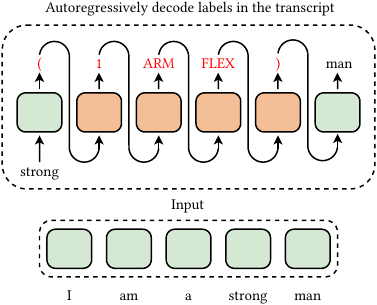}
  \caption{The overview of generative semantic gesture retrieval.}
  \label{fig:generative_retrieval}
\end{figure}

In this section, we introduce Generative Semantic Gesture Retrieval. The traditional index-retrieval-rank paradigm \cite{LLM4IR} retrieves candidate items from a database based on the context of the input query and then ranks these items to determine the final item. Although this method thoroughly considers the semantic distribution of the database, it only supports whole-sentence retrieval and cannot automatically determine the specific occurrence timestamp of the retrieved item within the sentence. To solve it, we choose the generative retrieval architecture inspired by GENRE \cite{GENRE}. Specifically, as shown in \fig\ref{fig:generative_retrieval}, the whole process is modeled as the standard prompt-based autoregressive generation. Given the input transcript as the prompt, the model will repeat the input and, based on the context, insert the retrieved semantic gesture, e.g., \gesLabel{1 ARM FLEX}, at the appropriate position, where \gesLabel{1} represents the index of the gesture. The retrieval model is initialized by the Large Language Model (LLM), ensuring its generalizability. Then, it is fine-tuned with a small amount of annotated data for task alignment and to perceive the semantic distribution of the semantic gesture set, with specific details to be discussed in Section \ref{subsec:llm_retrieval_model}.

\begin{figure}[t]
  \centering
  \includegraphics[width=.9\linewidth]{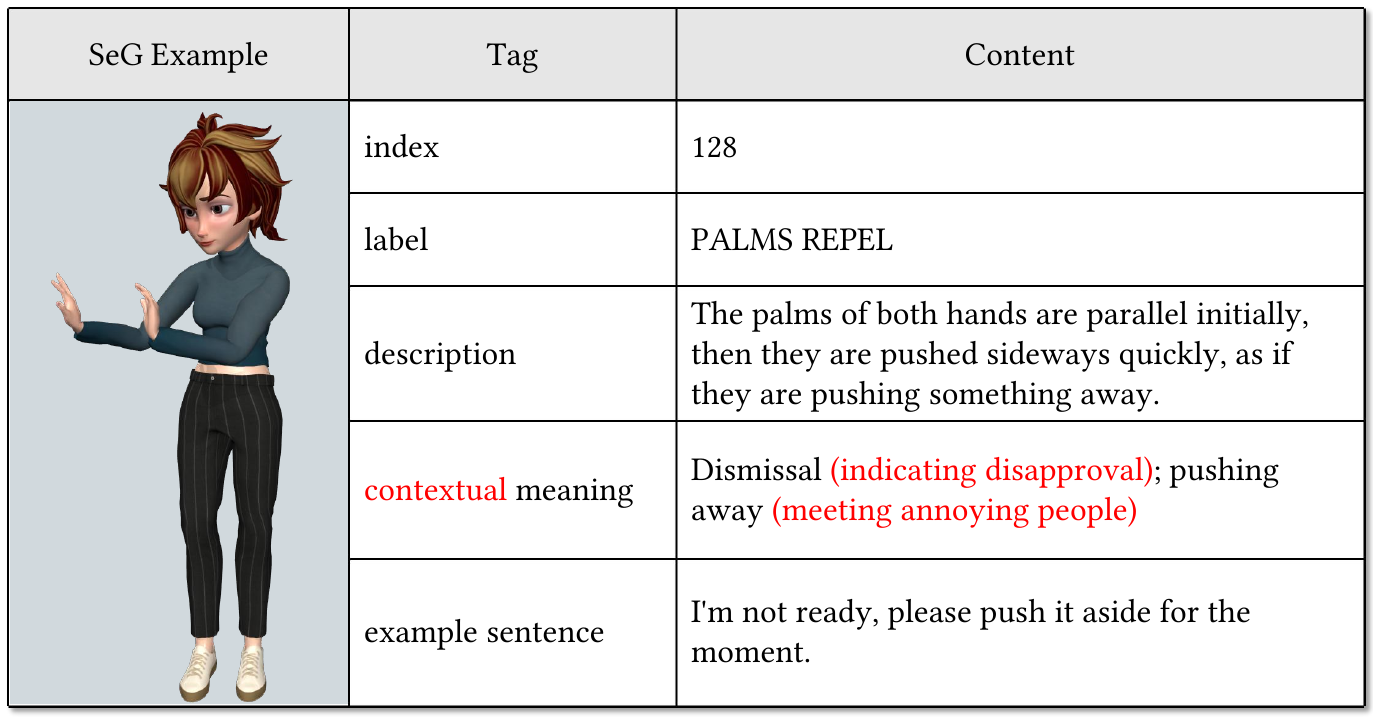}
  \caption{The meta-information of a semantic gesture in the SeG Dataset.}
  \label{fig:seg_example}
\end{figure}

\begin{figure}[t]
  \centering
  \includegraphics[width=.8\linewidth]{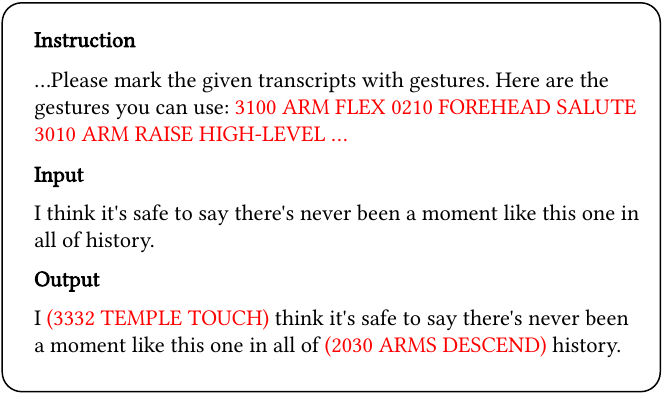}
  \caption{An example of the instruction dataset for (Large Language Model) LLM fine-tuning.}
  \label{fig:instruction_data}
\end{figure}

\subsection{SeG: Semantic Gesture Dataset}
\label{subsec:semantic_ges_dataset}
Gestures are intrinsically linked with semantic information and form a fundamental aspect of human communication. This connection dates back to the use of body postures for expressing intentions, predating the use of written symbols. Meanwhile, the semantic information conveyed by gestures is significantly influenced by cultural backgrounds, leading to substantial differences in gesture meanings across various cultures. To cover such diversity and avoid the ambiguity inherent in semantic gestures, we reference and synthesize findings in the fields of linguistic and human behavioral studies pertaining to semantic gestures \cite{Morris1994BodytalkAW,Wagner2003FieldGT,Kipp2005GestureGB,GESTUNO}. This effort leads to the compilation of a comprehensive semantic gesture dataset (SeG), denoted as $\boldsymbol{G} = [\vect{g}_i]_{i=1}^{N_{\eqword{sec}}}$, which encompasses $N_{\eqword{sec}}$ types of semantic gestures commonly utilized globally. Each record $\vect{g}$ in the dataset consists of a sequence of captured motion and its meta-information. \fig\ref{fig:seg_example} shows an example of the meta-information of a semantic gesture in SeG, which includes an index, label, description, contextual meaning, and example sentence, facilitating a thorough organization and analysis of human gestures.

As for motion, we employ professional motion capture equipment to gather high-quality, finger-contained animation. The collection involves two performers, a male and a female. To ensure the diversity of motion, performers deliver multiple interpretations and styles for each semantic gesture in the dataset. In total, we capture 1.5 hours of semantic gesture motion data, with each gesture being interpreted in an average of 5.7 in different ways.

\subsection{LLM-Based Retrieval Model}
\label{subsec:llm_retrieval_model}

Concurrently, we collect twenty 10$\sim$20-minute speech transcripts from TED talks~\cite{TED_talks} and enlist annotators for manual annotation, thereby creating an instruction dataset to fine-tune Large Language Models (LLMs) for semantic gesture retrieval. \fig\ref{fig:instruction_data} shows a case of the instruction dataset. Notably, we explicitly integrate the index information of each gesture into the \gesLabel{Instruction} part of the dataset and require the model to predict the name of the gesture as well as its index simultaneously. In practice, this trick could significantly alleviate the model to produce hallucinations, including the generation of non-existent gestures and the combination of different body parts and other actions. Intuitively, incorporating such logical information can help enhance the reasoning ability of LLMs, similar to how some studies opt to add some code data when training LLMs \cite{LLaMA,anil2023palm2}. We choose to fine-tune OpenAI's GPT-3.5-turbo~\cite{GPT3.5} on the instruction dataset.

\begin{figure}[t]
  \centering
  \includegraphics[width=\linewidth]{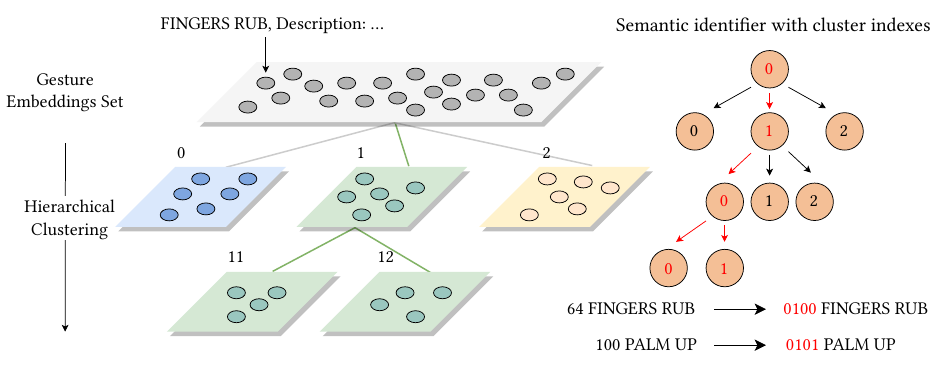}
  \caption{The process of building a semantics-aware identifier for the semantic gesture.}
  \label{fig:indexing_identifier}
\end{figure}

Meanwhile, the current instruction data does not take into account the meta-information of semantic gestures, thereby overlooking the rich semantic information it contains. In addition to using \gesLabel{example sentence} from the meta-information of each gesture to enrich the dataset, we also replace the current indexing identifier for each gesture, a simple combination of increasing numbers and gesture names, with a new semantics-aware indexing identifier. The core idea is to cluster the semantic gestures according to their textual description and index them based on the clusters, making the index implicitly indicate the relationship between different gestures. To be specific, as illustrated in \fig\ref{fig:indexing_identifier}, we feed the gesture \gesLabel{label}, \gesLabel{description}, and \gesLabel{contextual meaning} into a sentence-T5-base model \cite{T5} to generate a representational embedding vector. Each vector encapsulates the semantic information of a specific gesture within the SeG, and collectively, these vectors form a big gesture embedding set. Then, we apply hierarchical clustering with constrained K-means~\cite{bennett2000constrained} to these embedding vectors. The resultant index within each cluster is utilized as the final hierarchical semantic number identifiers. We conduct a related ablation study in Section \ref{subsubsec:semantics-aware_indexing_identifier}.

\begin{figure}[t]
  \centering
  \includegraphics[width=.9\linewidth]{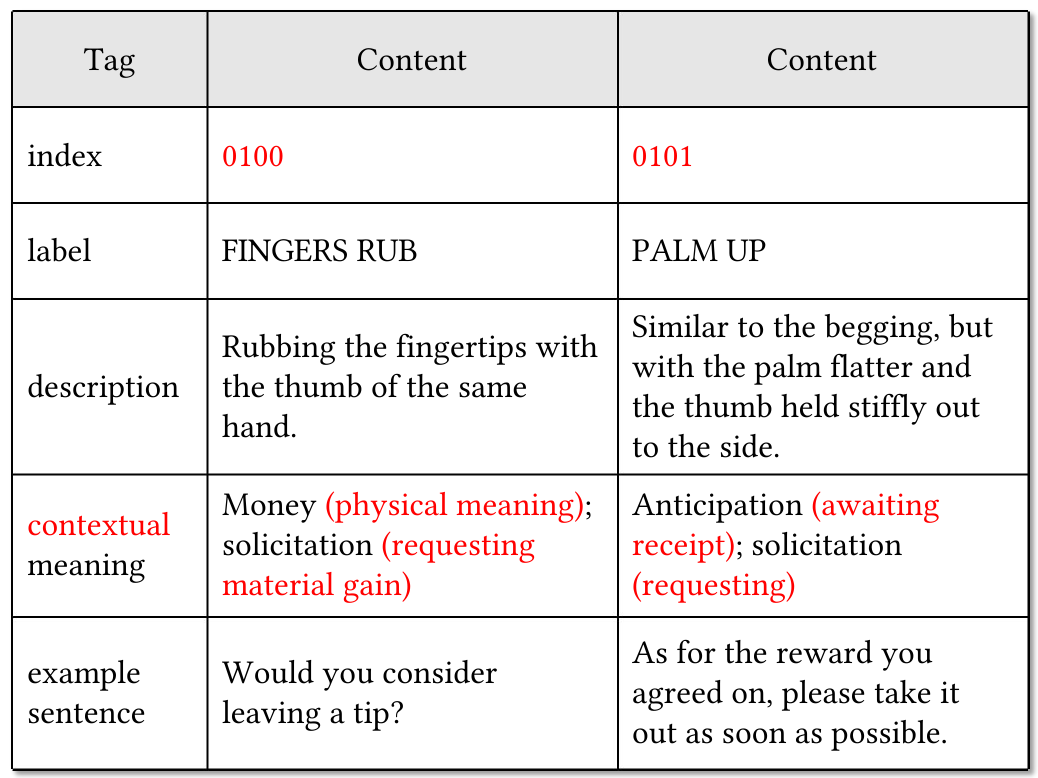}
  \caption{Comparison of the indexing results of two semantic gestures.}
  \label{fig:identifier_example}
\end{figure}

For example, as shown in \fig\ref{fig:identifier_example}, the gestures \gesLabel{FINGERS RUB} and \gesLabel{PALM UP} share the same prefix "010" and differ in the $4$-th position, indicating similar description or contextual meanings. With these indexes, the final semantic identifiers, e.g., \gesLabel{0100 FINGERS RUB}, can capture hierarchical semantics of gestures while ensuring interpretability.
\section{Semantics Gesture Alignment}
\label{sec:alignment}
We develop an alignment module to merge retrieved semantic gestures into rhythmic gestures synthesized by the generator $\mathcal{G}$ at the right timing. Specifically, we need to solve two problems: (a) determining \emph{when} to merge each semantic gesture; and (b) \emph{how} to merge these semantic gestures into the generated rhythmic gestures without affecting the naturalness of the original motion. Corresponding solutions are discussed in the following sections.

\begin{figure}[t]
  \centering
  \includegraphics[width=\linewidth]{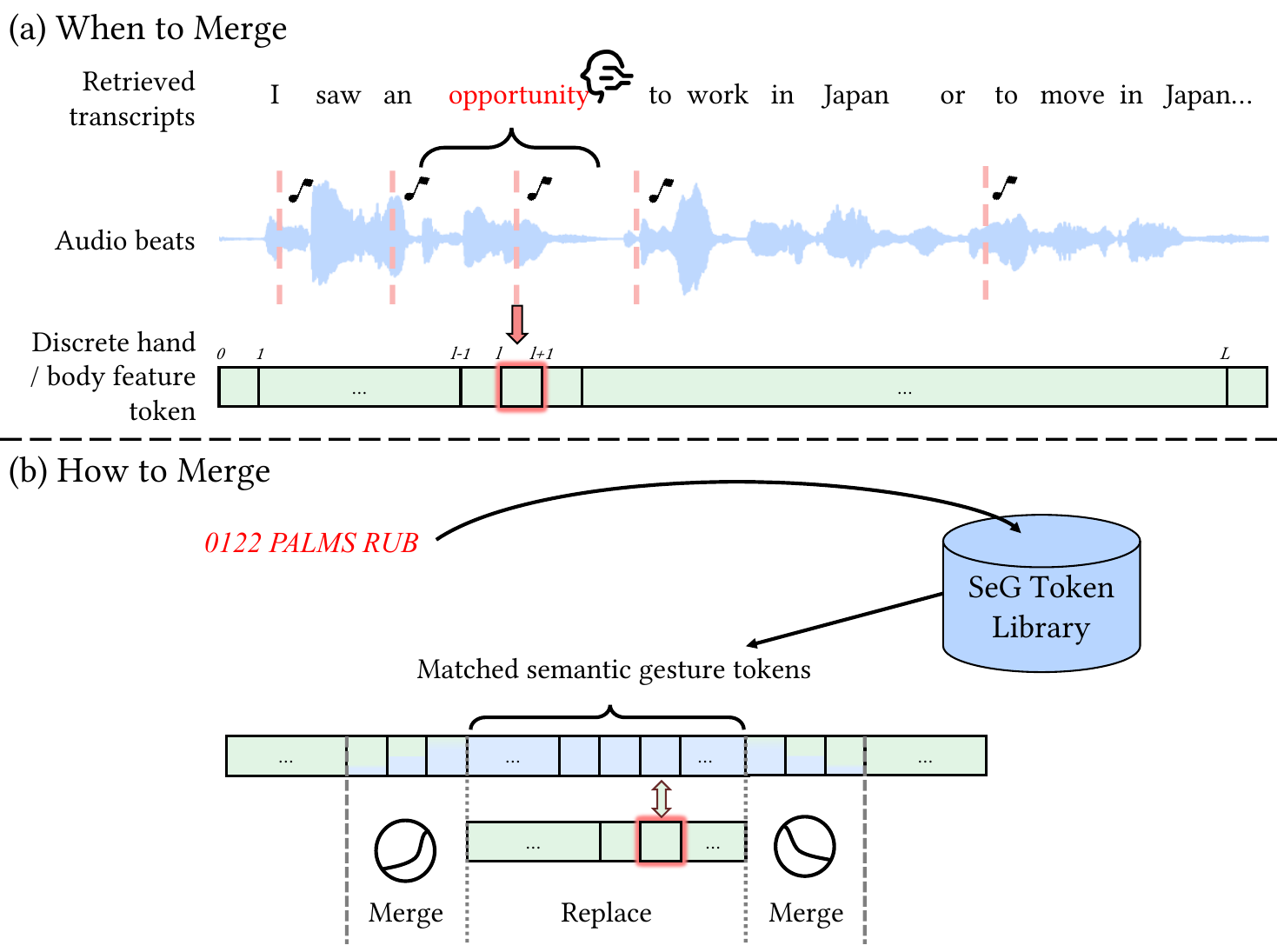}
  \caption{(a) Determining \emph{when} to merge each semantic gesture based on audio beats; and (b) \emph{how} to merge these semantic gestures into the generated rhythmic gestures without affecting the naturalness of the original motion.}
  \label{fig:align}
\end{figure}

\subsection{When to Merge}
\label{subsec:alignment_timestamp}
To determine when to merge retrieved semantic gestures, a straightforward method is to directly mark the timestamp of the trigger word (such as "opportunity" in \fig{\ref{fig:align}} (a)) identified by the retrieval model as the \emph{merging timing} $l$. The trigger word timestamp refers to the midpoint of the time interval in which the word occurs. But this approach would disrupt the ongoing beat gestures or create gestures with unexpected strokes during the subdued parts of the audio, leading to rhythm incongruity. To solve it, as shown in \fig{\ref{fig:align}} (a), our approach begins with detecting beats of the audio, which likely correspond to the stroke phase of gestures. This phase is recognized as the most expressive and energetically concentrated phase within a gesture's progression ~\cite{FERSTL2020117}. Subsequently, we identify the beat closest to the timestamp of the trigger word as the final $l$. The positive impact of this approach on rhythm integrity will be discussed in an ablation study.

\subsection{How to Merge}
\label{subsec:replacement}
In this section, we aim at explicitly merging retrieved semantic gestures with generated rhythmic gestures at the merging timing $l$. For preparation, we first extract the stroke part of each semantic gesture from the SeG Dataset and crop it into a $1$-second segment. Then, these gesture segments are encoded into discrete tokens using $\mathcal{E}_{\eqword{VQ}}$, thereby constructing the SeG token library. Given the index and label of the retrieved semantic gesture, we could search the matched gesture tokens from the library. As illustrated in \fig{\ref{fig:align}} (b), the following merging process is executed in two phases: (a) we replace the original motion with matched semantic gesture tokens, while aligning the timing $l$ with the three-quarters of the way through the matched token sequences. It allows semantic gestures to manifest slightly ahead of the corresponding semantic content in the speech, reflecting the typical $0.4$-second planning phase observed in human gestures informed by the behavioral study findings by ~\cite{INDEFREY2004101}. To mitigate abrupt and unnatural transitions often associated with hard replacements, inspired by TM2D ~\cite{gong2023tm2d}, we employ a strategy involving a weighted merge operation at token positions around the replacement site. Explicitly, we increase the merging weight of the matched semantic gesture tokens $w_{s}$ from 0.3 to 0.7 using a half cosine curve prior to replacement, and subsequently reduce it in the opposite manner after replacement. Meanwhile, the merging weight of the raw motion tokens $w_{r}$ is set to 1-$w_{s}$ to maintain the scale of the features. This technique ensures the final animation decoded by $\mathcal{D}_{\eqword{VQ}}$ results is not only smooth but also retains the intended semantic richness and rhythmic precision. 

Additionally, the gesture generator $\mathcal{G}$, during pre-training, only utilizes paired speech-gesture data and has not encountered the semantic gesture tokens from the SeG dataset. Before merging, it is necessary to ensure the output motion distribution of $\mathcal{G}$ covering the SeG dataset. Inspired by the Supervised Fine-Tuning (SFT) strategy for LLMs \cite{Selfinstruct}, we utilize the speech transcripts with semantic annotation in Section \ref{subsec:llm_retrieval_model} and the alignment timestamps of each semantic gestures to construct a new instruction dataset to fine-tune the pre-trained generator $\mathcal{G}$. The instruction dataset consists of the speech audio synthesized by a Text-to-Speech (TTS) tool \cite{murfai2022tts}, semantic gesture tokens encoded by $\mathcal{E}_{\eqword{VQ}}$, and their alignment timestamps. When fine-tuning, the generator takes the synthesized speech audio as input and generates the synchronized gestures, where we only optimize the cross-entropy loss at the merging interval for the target semantic gesture tokens. It could implicitly align the generator with the distribution of semantic gestures and make a preparation for the subsequent explicit merging.

\section{Evaluation}
\label{sec:results}
In this section, we initially outline our system's setup, followed by an evaluation of our results. We then compare these results with those from other systems, discuss potential applications, and validate our framework's various design choices through an ablation study.

\begin{figure*}[t]
    \centering
    \includegraphics[width=\linewidth]{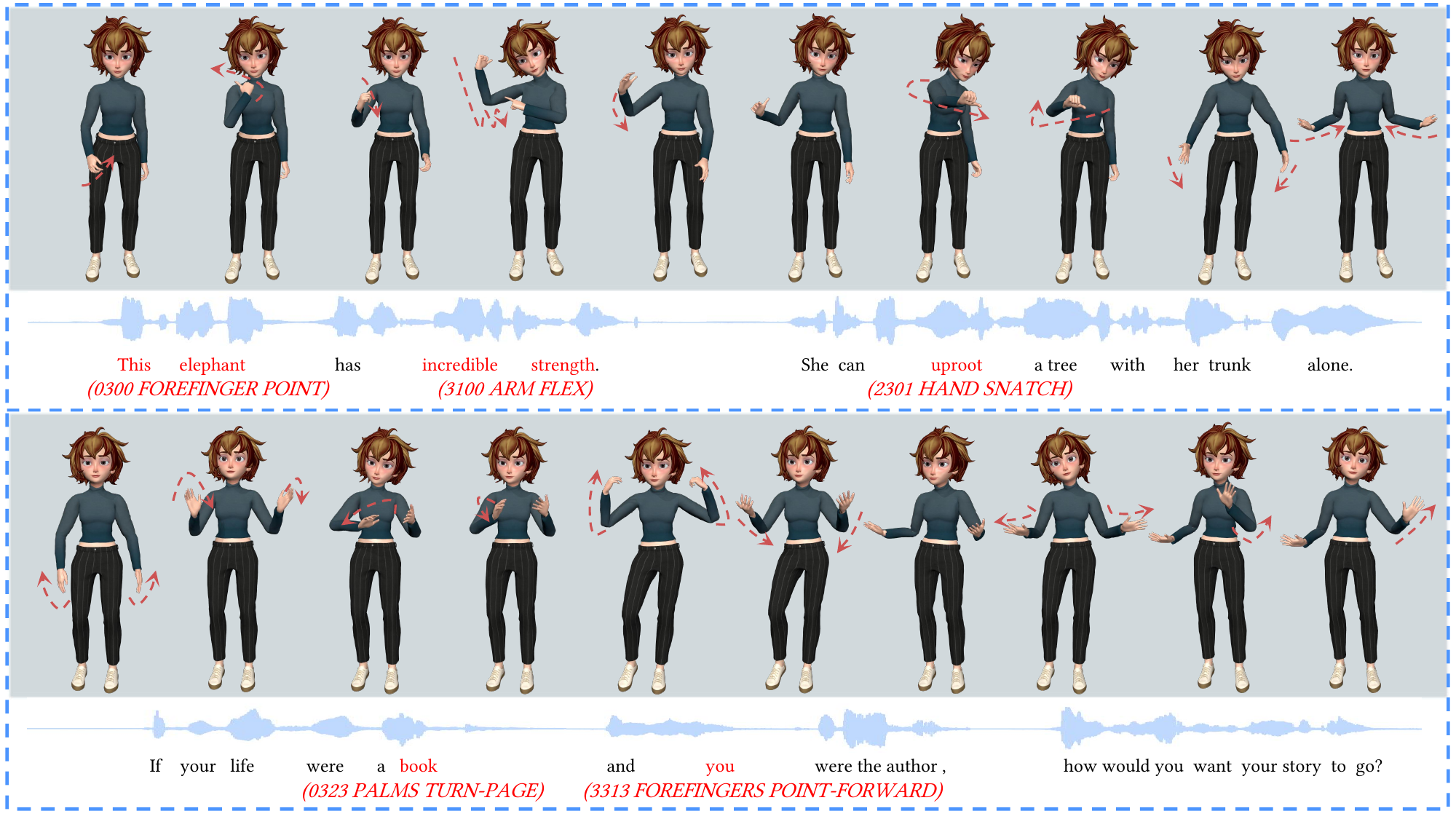}
    \caption{Semantics-aware gestures synthesized by our system conditioned on results of the semantic gesture retrieval and the synchronized speech. The character performs a variety of semantic gestures, including significant body movements, reasonable arm swings, and delicate finger gesticulations.}
    \Description{}
    \label{fig:qualitative_result}
\end{figure*}

\begin{figure*}[t]
    \centering
    \includegraphics[width=\linewidth]{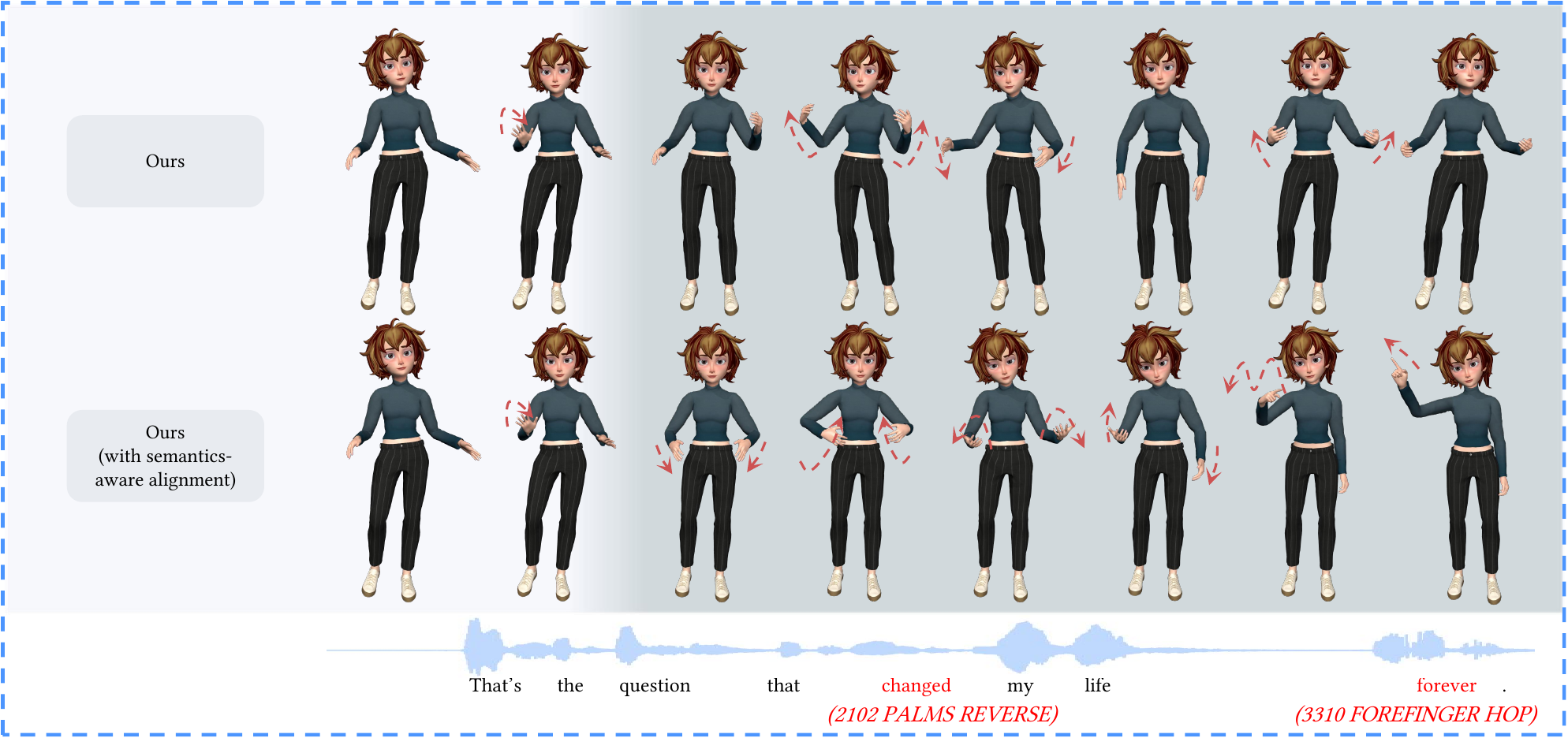}
    \caption{Qualitative comparison between synthesis without semantics-aware alignment (first row) and synthesis with alignment (second row).}
    \Description{}
    \label{fig:ablation_semantic_alignment}
\end{figure*}

\subsection{System Setup}
\label{subsec:system_setup}
\subsubsection{Speech-Gesture Datasets}
As for the speech-gesture datasets, our system is trained and evaluated on two high-quality speech-gesture datasets: \emph{ZEGGS} \cite{ghorbani2022zeroeggs} and \emph{BEAT} \cite{liu2021beatdataset}. The ZEGGS Dataset comprises two hours of full-body motion capture and audio from monologues by an English-speaking female actor, performed in 19 distinct styles. Synchronized transcripts are obtained using an automatic speech recognition (ASR) tool \cite{alibaba2009asr}. The BEAT dataset features about 76 hours of multimodal speech data, including audio, transcripts, and full-body motion capture from 30 speakers. These speakers performed in eight emotional styles across four languages. In alignment with the baseline setting in \cite{liu2021beatdataset}, we selectively use the speech data of English speakers 2, 4, 6, and 8. 

\subsubsection{Instruction Dataset for LLM Fine-Tuning}
To construct the annotation instruction dataset in Section \ref{subsec:llm_retrieval_model}, we collect twenty 10$\sim$20-minute speech transcripts and annotate them manually. For gestures not annotated in this process, we use examples from SeG dataset and employ few-shot strategy to prompt GPT-4 \cite{openai2023gpt4} in generating new annotated sentences containing them. This process mitigates the performance decline caused by the long-tail distribution of semantic gestures. When fine-tuning the LLM, there are a total of 5,410,940 training tokens for naive indexing identifier and 6,113,270 training tokens for semantics-aware indexing identifier, respectively.

\subsubsection{Settings}
Our system generates semantic-aware gestures at a rate of $60$ fps. In configuring the RVQ (Section \ref{subsec:gesture_vq-vae}), we set the number of residual quantization layers to $R = 4$, the downsampling rate to $d = 8$, and the codebook size to $C = 512$. This system is trained on both the SeG dataset and the speech-gesture datasets (Section \ref{subsec:system_setup}). The parameters $w_1$, $w_2$, $w_3$, and $w_4$ in the loss function are assigned values of $1$, $1$, $0.02$, and $1$, respectively. Regarding the Gesture Generator $\mathcal{G}$ (Section \ref{subsec:gesture_generator}), it is composed of a $12$-layer transformer with a width of $768$ features and is trained exclusively on speech-gesture datasets. During training, the RVQ is trained using standard motion clips of $2$ seconds in length, whereas $\mathcal{G}$ is trained with paired speech and motion clips of $4$ seconds. All the models are trained using four NVIDIA 3090Ti GPUs, taking approximately two days for the ZEGGS dataset and five days for the BEAT dataset. For LLM fine-tuning, we train $5$ epochs and set batch size to $2$, taking approximately one hour.

\subsection{Results}
\fig\ref{fig:qualitative_result} shows the visualization results of our system generating gestures conditioned on the semantic gesture retrieval outcomes and synchronized speech. The generator $\mathcal{G}$ employs top-5 sampling methods for token generation during inference and we utilize a physical tracking approach from \cite{yao2023moconvq} to alleviate the issue of sliding foot in our results. Our system successfully creates realistic gestures that accurately convey the intended meanings, aligning with the respective retrieval results. The character performs a range of semantic gestures, including natural body movements, reasonable arm swings, and delicate finger gesticulations. For example, the character performs suitable pointing gestures when ``elephant" and ``you" are mentioned. In the meanwhile, ``uproot" and ``book" are expressively portrayed. For abstract concepts like "strength," the character vividly illustrates them by gesturing towards the biceps. In the supplementary video, when the word ``two" is mentioned, the character naturally makes a ``V-sign" gesture. This demonstrates that the retrieval model can accurately capture the specific meanings of each semantic gesture in the SeG dataset.

To illustrate the significance of the semantics-aware alignment module, \fig\ref{fig:ablation_semantic_alignment} presents the comparative visualization results of synthesis without semantics-aware alignment (the first row) against synthesis with such alignment (the second row). The communicative efficacy of the generated gestures is enhanced through the explicit integration of contextually appropriate, meaningful gestures based on the speech transcript.

\begin{table*}[t]
    \centering
    \caption{Average scores of user study with $95\%$ confidence intervals. \emph{Our system without semantic alignment (w/o semantic alignment)} excludes the semantics-aware alignment module for the pre-trained generator. Asterisks indicate the significant effects ($*: p < 0.05$, $**: p < 0.01$, $***: p < 0.001$).}
    \label{tab:user_study}

    \newcolumntype{Y}{>{\raggedleft\arraybackslash}X}
    \newcolumntype{Z}{>{\centering\arraybackslash}X}
    \begin{tabularx}{\linewidth}{llYYY}
        \toprule
        Dataset & System & Human Likeness $\uparrow$ & Beat Matching $\uparrow$ & Semantic Accuracy $\uparrow$ \\ 
        \toprule
        \multirow{4}*{ZEGGS} & GT & $0.07 \pm 0.02^{*}$ & $0.15 \pm 0.03^{*}$ & $0.51 \pm 0.08$ \\
        \cline{2-5}
        & GestureDiffuCLIP & $-0.02 \pm 0.01^{*}$ & $-0.05 \pm 0.01$ & $-0.15 \pm 0.10^{**}$ \\
        & Ours (w/o semantic alignment) & $-0.08 \pm 0.03^{*}$ & $-0.07 \pm 0.02$ & $-0.94 \pm 0.12^{**}$ \\
        & Ours & $0.02 \pm 0.01$ & $-0.05 \pm 0.02$ & $\bm{0.48 \pm 0.07}$ \\
        
        \midrule
        \multirow{4}*{BEAT} & GT & $0.43 \pm 0.06$ & $0.39 \pm 0.06$ & $0.37 \pm 0.05$ \\
        \cline{2-5}
        & CaMN & $-1.03 \pm 0.15^{**}$ & $-0.91 \pm 0.13^{**}$ & $-0.22 \pm 0.04^{**}$ \\
        & Ours (w/o semantic alignment) & $0.29 \pm 0.08$ & $0.32 \pm 0.07$ & $-0.58 \pm 0.07^{**}$ \\
        & Ours & $\bm{0.35 \pm 0.04}$ & $\bm{0.33 \pm 0.05}$ & $\bm{0.41 \pm 0.03}$ \\
        \bottomrule
    \end{tabularx}

\end{table*}

\begin{figure*}[t]
    \centering
    \includegraphics[width=\linewidth]{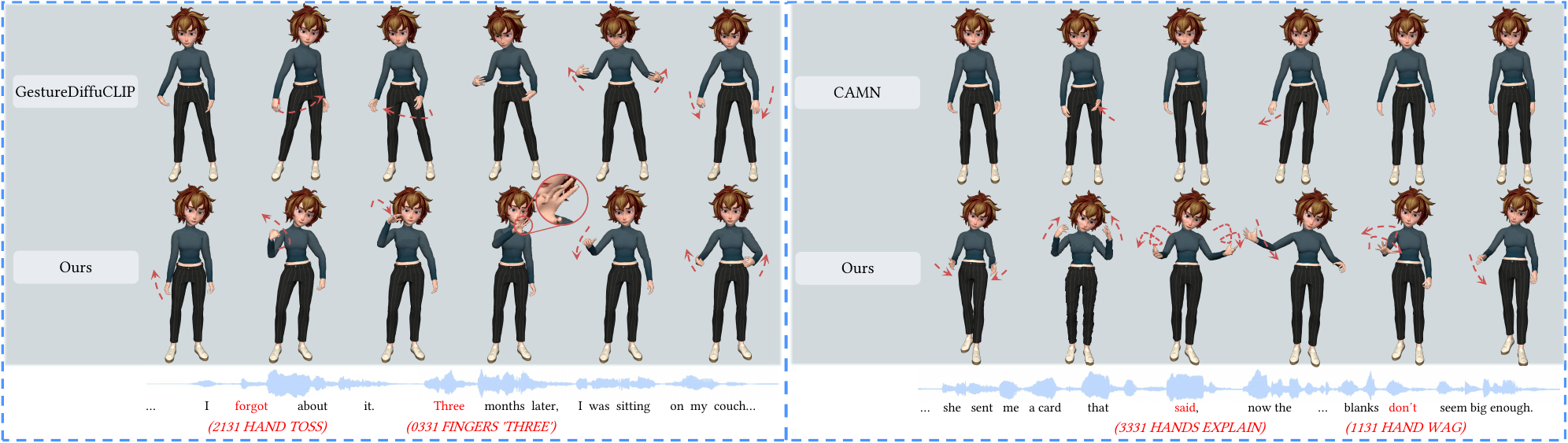}
    \caption{Qualitative comparison between our system and baselines (GestureDiffuCLIP \cite{ao2023gesturediffuclip} and CaMN \cite{liu2021beatdataset}) using two test speech excerpts.}
    \Description{}
    \label{fig:comparison}
\end{figure*}

\subsection{Comparison}
Evaluating gestures with objective metrics presents significant challenges. Many current objective metrics poorly correlate with subjective feedback outcomes~\cite{kucherenko2023evaluating}. Echoing approaches in \cite{alexanderson2023listendenoiseaction,ghorbani2022zeroeggs,ao2023gesturediffuclip}, this study emphasizes user study evaluations for generated results, with quantitative evaluations serving as supplementary references.

\subsubsection{Baselines}
We compare our system with GestureDiffuCLIP~\cite{ao2023gesturediffuclip} on the ZEGGS dataset, using the code provided by its authors. This system learns an implicit shared space between transcripts and gestures, enhancing semantic perception. The BEAT dataset is released alongside a strong baseline, the Cascaded Motion Network (CaMN), which uses transcripts as inputs to generate semantic gestures based on a hierarchical architecture. The related codes for two semantics-aware systems~\cite{zhi2023livelyspeaker,gao2023gesgpt} are not available at the time of writing this work.

\subsubsection{User Study}
\label{subsubsec:user_study}
Following the method in \cite{alexanderson2023listendenoiseaction,ao2023gesturediffuclip}, we conduct user studies using pairwise comparisons. For each test, participants view two 10-second videos, each synthesized by different models (including the ground truth) for the same speech segment, played one after the other. Participants are required to choose the video they prefer, following the instructions given below the videos, and rate their choice on a scale from $0$ to $2$, where $0$ signifies no preference. The unselected video in the pair is then assigned the inverse score (for instance, if the chosen video is rated $1$, the other video is assigned $-1$). Participant recruitment is conducted via the Credamo platform \cite{credamo}. Details of the user study are described in Appendix \ref{sec:details_of_user_study}.

We conduct three distinct preference tests: \emph{human likeness}, \emph{beat matching}, and \emph{semantic accuracy}, incorporating attention checks in each. During the \emph{human likeness} test, participants determine if the generated motion closely mimics that of a real human. To avoid any speech-induced bias, these video clips are presented without sound. In the \emph{beat matching} test, participants assess the synchronization of the generated motion with the speech's rhythm. For the \emph{semantic accuracy} test, participants are required to evaluate whether the generated gestures based on the input speech accurately convey the appropriate semantics. The average scores from these tests are detailed in Table \ref{tab:user_study}. We implement a one-way ANOVA and a post-hoc Tukey multiple comparison test for each user study. The assumptions of normality, homogeneity of variances, and independence for each ANOVA are all met for both the ZEGGS and BEAT datasets.

In the ZEGGS Dataset evaluation, we analyze four methods: the ground-truth gestures (GT), our system (Ours), our system without the semantic alignment module (w/o semantic alignment) for ablation, and GestureDiffuCLIP. After attention checks, we collect the valid answers of $98$, $99$, $101$ participants for the human likeness, beat matching, and semantic accuracy tests, respectively. Multiple one-way ANOVAs are conducted, one for each questionnaire item. The results reveal that different baselines have statistically significant effects on 
\emph{human likeness} ($F(3, 4700) = 142.91$, $p < .05$, $\eqword{Partial Eta Squared} = 0.083$), 
\emph{beat matching} ($F(3, 4748) = 119.71$, $p < .05$, $\eqword{Partial Eta Squared} = 0.071$), 
and \emph{semantic accuracy} ($F(3, 4844) = 250.87$, $p < .01$, $\eqword{Partial Eta Squared} = 0.134$).
Table \ref{tab:user_study} indicates that, in the human likeness and beat matching tests, the performance differences between Ours and the other methods are not significant. But in the semantic accuracy test, Ours notably excels over both GestureDiffuCLIP and Ours w/o semantic alignment, with a considerable margin ($p < 0.001$), emphasizing the vital role of the semantic alignment module in enhancing semantic perception. The left part of \fig\ref{fig:comparison} provides a visual demonstration, showing that the motions generated by GestureDiffuCLIP are less meaningful compared to those from Ours. These results confirm the efficiency of our system in semantic gesture synthesis.

On the BEAT Dataset, our evaluation encompasses four methods: the ground-truth gestures (GT), our system (Ours), our system without the semantic alignment module (w/o semantic alignment) for ablation, and CaMN, with its speaker ID input matched to the ground truth. In the user study, $99$, $103$, and $100$ subjects pass the attention checks for the human likeness, beat matching, and semantic accuracy tests, respectively. Multiple one-way ANOVAs indicate that different generation methods have main effects on 
\emph{human likeness} ($F(3, 4748) = 179.92$, $p < .01$, $\eqword{Partial Eta Squared} = 0.102$), 
\emph{beat matching} ($F(3, 4940) = 271.35$, $p < .01$, $\eqword{Partial Eta Squared} = 0.141$), 
and \emph{semantic accuracy} ($F(3, 4796) = 151.52$, $p < .01$, $\eqword{Partial}$ $\eqword{Eta}$ $\eqword{Squared} = 0.087$).
As shown in Table \ref{tab:user_study}, in the human likeness and beat matching tests, GT, Ours, and Ours w/o semantic alignment perform comparably and surpass CaMN ($p < 0.001$). For the semantic accuracy test, Ours outscore other baselines ($p < 0.001$), but the score of Ours (w/o semantic alignment) decreases significantly due to the lack of semantic alignment. It highlights the essential role of the semantic alignment module in maintaining semantic consistency between speech and gestures. The right part of \fig\ref{fig:comparison} illustrates that gestures generated by our system exhibit greater communicative efficacy compared to those produced by CaMN.

\begin{table*}[t]
    \centering
    \caption{Quantitative evaluation on the ZEGGS and BEAT Datasets. This table reports the mean ($\pm$ standard deviation) values for each metric by synthesizing on the test data $10$ times.}
    \label{tab:quantitative_evaluation}

    \newcolumntype{Y}{>{\raggedleft\arraybackslash}X}
    \newcolumntype{Z}{>{\centering\arraybackslash}X}
    \begin{tabularx}{\linewidth}{llYY}
        \toprule
        Dataset & System & FGD $\downarrow$ & SC $\uparrow$ \\ 
        \toprule
        \multirow{4}*{ZEGGS} & GT & - & $0.55$ \\
        \cline{2-4}
        & GestureDiffuCLIP & $81.73 \pm 3.27$ & $0.21 \pm 0.07$ \\
        & Ours (w/o semantic alignment) & $82.02 \pm 2.49$ & $0.09 \pm 0.02$ \\
        & Ours (w/ naive indexer) & $81.88 \pm 2.23$ & $0.30 \pm 0.03$ \\
        & Ours & $\bm{81.22 \pm 2.53}$ & $\bm{0.38 \pm 0.05}$ \\
        
        \midrule
        \multirow{4}*{BEAT} & GT & - & $0.65$ \\
        & CaMN & $105.42 \pm 0.00$ & $0.21 \pm 0.00$ \\
        & Ours (w/o semantic alignment) & $89.73 \pm 2.11$ & $0.08 \pm 0.02$ \\
        & Ours (w/ naive indexer) & $89.58 \pm 2.23$ & $0.38 \pm 0.01$ \\
        & Ours & $\bm{89.15 \pm 2.06}$ & $\bm{0.45 \pm 0.09}$ \\
        \bottomrule
    \end{tabularx}
\end{table*}

\subsubsection{Quantitative Evaluation}
\label{subsec:quantitative_evaluation}
We quantitatively evaluate the human likeness of generated motion and speech-gesture semantic matching using two metrics, i.e., Fr{\'e}chet Gesture Distance (FGD) \cite{yoon2020trimodalgesture} and Semantic Score (SC) \cite{ao2023gesturediffuclip}, respectively. The Fréchet Gesture Distance (FGD) quantifies the disparity between the latent feature distributions of generated and real gestures. Commonly employed to evaluate gesture perceptual quality, a lower FGD indicates superior motion quality. The Semantic Score (SC) assesses the semantic coherence between speech and generated gestures. It computes the cosine similarity in the semantic space between generated motion and ground-truth transcripts, as defined in the gesture-transcript embedding framework by \citet{ao2023gesturediffuclip}. SC ranges from -1 to 1, with a higher SC indicating more effective speech-gesture content alignment. Besides, the FGD and SC are computed using sentence-level motion segments. We compute the mean ($\pm$ standard deviation) values for each metric by generating on the test data 10 times.

As shown in Table \ref{tab:quantitative_evaluation},  our system surpasses all baseline comparisons in both metrics, FGD and SC. Notably, discarding the semantic alignment module leads to a substantial decrease in the SC value of our system, underscoring the module's critical role. Meanwhile, the FGD value remains relatively unchanged with the addition of the semantic alignment module. This suggests that the fusion operation within this module does not degrade the quality of the motion. The SC metric exhibits a decline in our system when semantic gestures are indexed with increasing numbers (w/ naive indexer). This trend suggests that the semantics-aware indexing identifier, as detailed in Section \ref{subsec:llm_retrieval_model}, can effectively enhance the performance of the gesture retrieval model.

\subsection{Gesture Editing}
\label{subsec:application}
\begin{figure}[t]
  \centering
  \includegraphics[width=\linewidth]{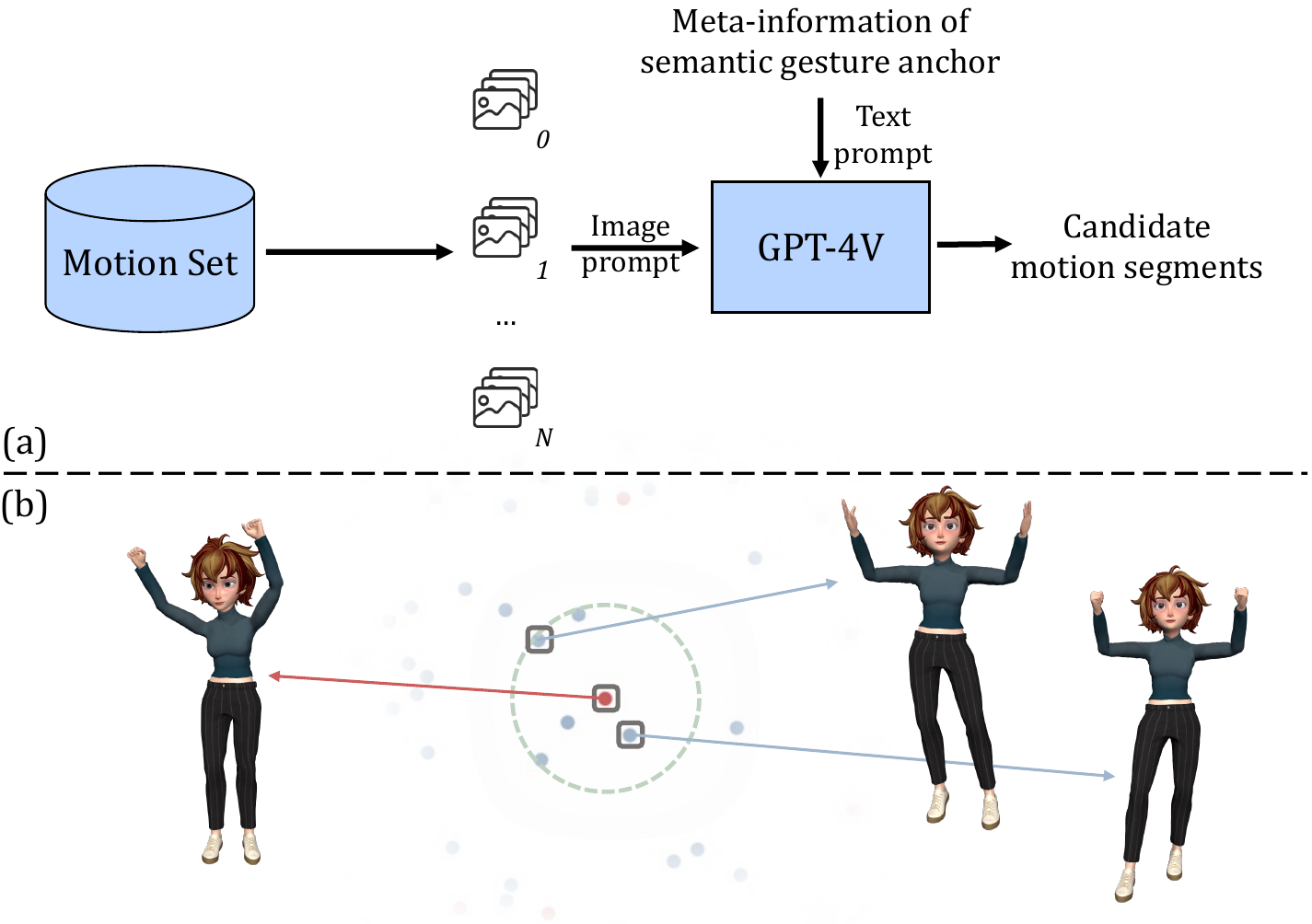}
  \caption{The data augmentation consists of two stages: (a) a high-level filtering utilizes the GPT-4V \cite{openai2024gpt4vision} to filter out candidate gesture segments from the motion base according to the meta-information of the semantic gesture anchor; and (b) an example of finding joint-level similar motion in the latent space of RVQ. The red point represents the embedding of semantic gesture \gesLabel{3030 ARMS RAISE V-SHAPE}. We can find similar motion within an appropriate threshold.}
  \label{fig:seg_augmentation}
\end{figure}

\begin{figure*}[t]
    \centering
    \includegraphics[width=\linewidth]{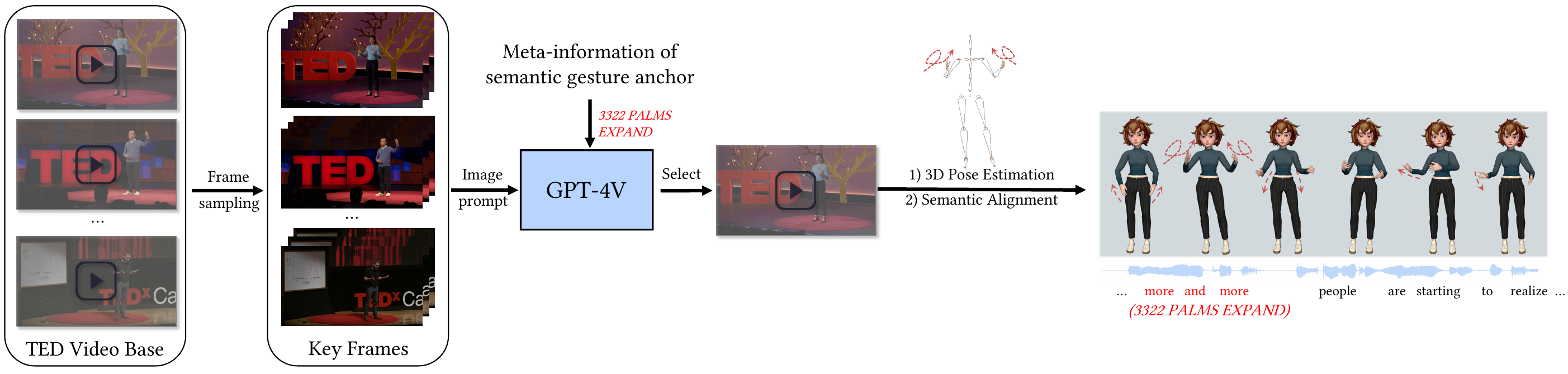}
    \caption{An example of gesture editing (Section \ref{subsec:application}). We use GPT-4V to filter appropriate semantic gestures from in-the-wild 2D videos to edit the original motion segment.}
    \Description{}
    \label{fig:aplication}
\end{figure*}

Our SeG dataset (Section \ref{subsec:semantic_ges_dataset}) faces two primary issues: a) the variations of each semantic gesture is restricted; and b) there is a discrepancy between the captured motion and the real, spontaneous speech gestures. To enrich the diversity and bridge the gap, we propose a data augmentation framework to retrieve the semantically similar gesture segment $\vect{M}'$ corresponding to the given semantic gesture $\vect{g}$ from other existing datasets, e.g., $\vect{G}'$. $\vect{G}'$ is derived from 2D videos. We measure the semantic relevance between gesture sequences in two dimensions: (a) a high-level filtering based on the meta-information of $\vect{g}$; and (b) a low-level movement matching between $\vect{M}'$ and motion part $\vect{M}$ of $\vect{g}$. Details of these methods are introduced in the following sections.

Based on the augmentation method, we are able to achieve flexible editing of the generated semantic gestures. For instance, as illustrated in \fig\ref{fig:aplication}, we employ a large multi-modality model to select appropriate 
semantic gestures from in-the-wild 2D videos conditioned on the meta-information of the retrieved semantic gesture, and align the new semantic gestures with original motion sequence. Furthermore, users can flexibly control the style and appearance of the final generated gestures by customizing a 2D video library. Please refer to the supplementary video for more visualization results.

\subsubsection{High-Level Filtering.}
We first utilize the GPT-4 with vision (GPT-4V) \cite{openai2024gpt4vision}, which is a powerful multimodal model and supports both image and text inputs, to filter out candidate gesture segments from $\vect{G}'$ according to the meta-information of the anchor $\vect{g}$. Specifically, as shown in \fig\ref{fig:seg_augmentation} (a), each segment in the motion set $\vect{G}'$ corresponds to a sequence of image frames. These image frames are about $1$ second in duration, and we uniformly sample $5$ images as the image prompt for GPT-4V. Meanwhile, the meta-information of the semantic gesture anchor is added as the text prompt. GPT-4V analyzes these prompts and selects the final candidate segments, matching them with the textual description of the anchor $\vect{g}$. These selected segments are converted into 3D motion sequences using a pose estimation tool \cite{deepmotion_animate}.

\subsubsection{Low-Level Matching.}
To refine the results of GPT-4V and ensure motion plausibility, we further select gestures from candidate motion segments that are similar to the semantic gesture anchor at the movement level. Specifically, we use the RVQ encoder $\mathcal{E}_{\eqword{VQ}}$ to encode the motion part $\vect{M}$ of the anchor and one of the selected motion segment $\vect{M}'$ into corresponding latent sequences $\vect{Z}$ and $\vect{Z}'$, respectively. These sequences are then averaged along the temporal dimension into $\overlinevect{z}$ and $\overlinevect{z}'$, respectively. We compute the Euclidean distance between these two embeddings as the similarity score. If the score falls below the predefined threshold, $\vect{M}'$ is selected. Finally, we obtain semantically relevant gestures from $\vect{G}'$, thereby enriching the diversity of the gesture library. \fig\ref{fig:seg_augmentation} (b) illustrates an example of this process.

\subsection{Ablation Study}
Our analysis focuses on the effects of various architectures of the gesture tokenizer, the semantics-aware indexing identifier, and differing configurations of the semantics-aware alignment module on our system's performance. These findings are detailed in \fig\ref{fig:ablation_semantic_alignment}, Table \ref{tab:user_study}, Table \ref{tab:quantitative_evaluation}, \fig\ref{fig:compare}, and the supplementary video.

\subsubsection{LLM-Based Retrieval Model.}
\label{subsubsec:llm_based_retrieval_model}
\begin{figure}[htbp]
    \centering
    \includegraphics[width=\linewidth]{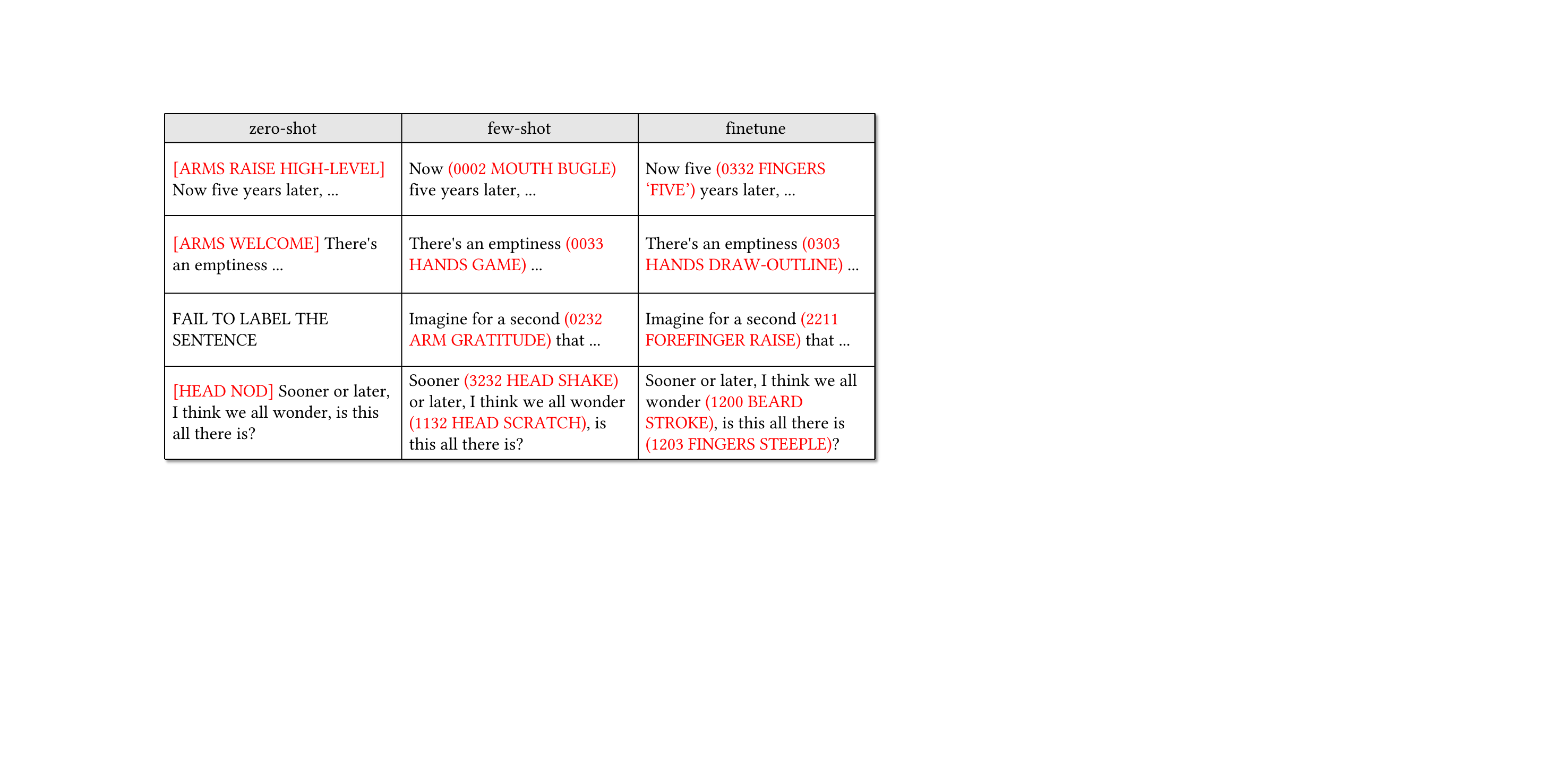}
    \caption{The comparison of three strategies.}
    \Description{}
    \label{fig:compare}
\end{figure}

In this experiment, we compare three different strategies, zero-shot, few-shot and finetuned LLM, which have exactly the same instruction settings and simultaneously contain the index information of all semantic gestures. Zero-shot strategy only has instruction settings, few-shot strategy has some examples of output formats, and the finetuned LLM is trained on the instruction dataset. We select GPT-3.5-turbo-1106 as the base model. There are two commonly used evaluation methods for LLM: automatic evaluation as quantitative analysis and human evaluation as qualitative analysis ~\cite{chang2023survey}. We conduct a qualitative analysis and present several examples in Figure \ref{fig:compare}. With the same instruction, the zero-shot LLM struggles to perform adequately, failing to generate valid gestures or comprehend semantic information effectively. For example, from Figure \ref{fig:compare} we can see that zero-shot LLM can not always produce appropriate gesture labels. The few-shot LLM effectively performs the annotation task and produces valid gesture labels but falls short in fully understanding semantic information. The fine-tuned model addresses all these requirements robustly, and can understand the semantics of gestures.

It is challenging to design an effective quantitative pipeline to evaluate the outcomes of LLM because there is a lack of credible metrics to judge whether the response is aligned with human performance. As explored in the NLP field ~\cite{zhang2023wider}, a potential approach is to utilize the LLM itself for evaluation and stabilize the results through multiple independent evaluations. Details and results of the quantitative evaluation are described in Appendix \ref{sec:details_of_llm_quantitative}.

\subsubsection{Architecture of Gesture Tokenizer.}
\label{subsubsed:architecture_of_gesture_tokenizer}
In this experiment, we train the gesture tokenizer (Section \ref{subsec:gesture_vq-vae}) with three different architectures, i.e., (1D-Conv+Transformer, RVQ), (1D-Conv, RVQ), and (1D-Conv+Transformer, VQ) to investigate the impact of the Transformer layer and the residual quantization layer on motion encoding, respectively. ``1D-Conv+Transformer" refers to a tokenizer whose encoder and decoder consist of 1D convolutional layers and the Transformer layer. We conduct comparisons by observing the quality of the reconstructed motion. As demonstrated in the supplementary video, motion reconstructed using RVQ based solely on 1D convolutional layers tend to exhibit jitter, while those using a vanilla quantization layer are prone to losing fine details, such as finger movements.

\subsubsection{Semantics-Aware Indexing Identifier.}
\label{subsubsec:semantics-aware_indexing_identifier}
In this experiment, we train and compare two generative retrieval models (Section \ref{subsec:llm_retrieval_model}) using the naive and semantics-aware indexing identifiers, respectively. The naive approach involves indexing semantic gestures with sequentially increasing numbers. Table \ref{tab:quantitative_evaluation} reveals that the semantics-aware metric, SC, experiences a decrease with the naive indexing approach. This observation confirms that the semantics-aware indexing identifier effectively improves the semantic consistency between speech and gestures.

\subsubsection{Semantics-Aware Alignment.}
We conduct three experiments to study the semantics-aware alignment module (Section \ref{sec:alignment}) from a holistic to a localized perspective. First, we discard the whole semantic alignment module, resulting in a significant drop in relevant semantic metrics, such as semantic accuracy (Table \ref{tab:user_study}) and SC (Table \ref{tab:quantitative_evaluation}). \fig\ref{fig:ablation_semantic_alignment} also demonstrates that gestures synthesized by Ours (w/o semantic alignment) perform low communicative efficacy. Then, we do not consider audio beats when determining the merging timing (Section \ref{subsec:alignment_timestamp}) for each retrieved semantic gesture. As shown in the supplementary video, the rhythmic harmony of the generated gestures is disrupted. Lastly, we discard the weighted merge operation (Section \ref{subsec:replacement}) when replacing the original motion with retrieved semantic gestures. As demonstrated in the supplementary video, the generated motion exhibits unnatural transitions.
\section{Conclusion}
\label{sec:conclusion}
In this paper, we present Semantic Gesticulator, a semantics-aware co-speech gesture synthesis system that generates 
realistic and meaningful gestures while maintaining rhythmic coherence with speech. Initially, we develop a GPT-based gesture generative model, trained using hierarchical discrete tokens extracted by a scalable, body part-aware residual VQ-VAE. This architecture effectively produces gestures that are rhythmically coherent and exhibit robust generalization across a wide range of audio inputs. Subsequently, we employ a powerful large language model (LLM) to establish a generative semantic gesture retrieval framework. This framework analyzes the context of speech transcripts and efficiently retrieves suitable semantic gestures from our self-collected, high-quality gesture library, namely the SeG Dataset. This dataset encompasses a comprehensive array of semantic gestures frequently utilized in human communication. Finally, we introduce a semantic alignment mechanism. This mechanism merges retrieved semantic gestures with generated rhythmic gestures at the latent space level, ensuring that the resulting gestures are both meaningful and rhythmically synchronized. We carry out a comprehensive series of experiments to assess our framework. Our system surpasses all baseline comparisons in terms of both qualitative and quantitative measures, demonstrated by the results of FGD and SC metrics, as well as user study outcomes. For applications, we devise an augmentation framework for identifying semantically similar gestures from an extensive collection of 2D videos, thereby enhancing the diversity of gestures. Moreover, through customizing a 2D video library, users can flexibly edit the style of final outcomes.

There is still room for improvement in our current research. 
First, the semantic gesture retrieval model only considers the textual information of each gesture and the input speech transcript, neglecting the inherent rhythm of speech audio. This may result in retrieving redundant gestures in areas lacking prominent prosody (moments when the speaker typically does not perform semantic gestures), or missing the corresponding semantic gestures in instances where prosody is significant. Employing a model to integrate multiple modalities, such as textual descriptions, gesture example images, and speech audio, can further enhance the accuracy of semantic gesture mining. Moreover, our retrieval model occasionally retrieves an excess of semantic gestures than necessary, failing to align with user preferences. We have observed that our annotators tend to label as many potential gestures as possible, which contributes to this issue. Our system currently supports users in manually revising the retrieval results. A potential way to achieve this automatically is to employ a LLM as a discriminator to refine the results of semantic gesture retrieval.A user may specify their preferences via prompting and let the LLM adjust the frequency of retrieved semantic gestures to a suitable degree.

Second, we employ a straightforward but effective merging strategy to align the retrieved semantic gestures with the generator's outcomes. However, we only align the stroke phase of semantic gestures to the audio beats to maintain the existing gesture sequence's phases. A more phase-informed strategy may better preserve motion details. Concurrently, augmenting the pool of semantic gesture samples with additional references and training a specialized generator could pave the way for the creation of a more varied range of semantic gestures. To further align the pre-trained generator $\mathcal{G}$ with semantic preferences or even human values, methods like reinforcement learning with human feedback \cite{christiano2023deep,GPT3.5} and Direct Preference Optimization (DPO)~\cite{DPO} can be explored to enhance the model's capabilities. 

Third, the motion quality exhibits certain deficiencies, such as foot sliding and excessive upper-body movements. The former issue can be alleviated by employing additional constraints in training and by using IK-based post-processing. The latter is partially caused by exaggerated gesture performance in motion data. Constructing a more curated dataset may address this problem.

Finally, our system enhances the communicative efficacy of generated gestures by explicitly specifying appropriate semantic gestures. Expanding this system to two-party, and even multi-party conversational scenarios, is a direction worth exploring in the future.

\begin{acks}
  We thank the anonymous reviewers for their constructive comments. This work was supported in part by National Key R\&D Program of China 2022ZD0160803.
\end{acks}

\bibliographystyle{ACM-Reference-Format}
\bibliography{gesture}

\clearpage

\appendix

\section{Details of User Study}
\label{sec:details_of_user_study}
A comparison set comprises two videos, each lasting 10 seconds, displayed sequentially from left to right. These pairs are generated using the same speech and character model. Questionnaires of the user study are built via the Human Behavior Online (HBO) tool provided by the Credamo platform \cite{credamo}. Each test and questionnaire consist of 24 such video pairs. On average, an experiment is completed in 12 minutes. We have recruited 220 participants through Credamo, of which 117 are male and 103 are female. 77 participants are 18 - 25 years of age, 121 are between 25 - 35, and 22 are above 35. They are sourced from the United States (30\%) and China (70\%) , and those involved in tests with audio must speak English fluently. In order to ensure that participants can clearly understand the detailed meaning of each metric and can distinguish between them effectively, We provide the participants with detailed instructions. For \emph{human likeness}, participants are requested to assess the naturalness and fluidity of the movements, as well as their resemblance to actual human movements. For \emph{beat matching}, we instruct the participants to focus on the rhythmic coherence between gestures and speech audio. For \emph{semantic accuracy}, the participants first learn about "what are semantic gestures" by watching several examples of semantic gestures. They are asked to read the transcript of each speech before rating, ensuring a more informed judgment on the semantic performance of the results. Meanwhile, to ensure the validity of the responses, following \cite{ao2023gesturediffuclip}, an attention check is incorporated randomly within the experiment. This check involves a text message, \emph{attention: please select the leftmost option}, displayed continuously at the video pair's bottom throughout the question and embedded in the video during the transition between the two clips. Responses failing this attention check are excluded from the final results.

The settings for the \emph{human likeness}, \emph{beat matching}, and \emph{semantic accuracy} tests on both the ZEGGS dataset and the BEAT dataset are identical. Specifically, We evaluate four methods for each dataset: GT, Ours, Ours (without semantic alignment), and either GestureDiffuCLIP for the ZEGGS dataset or CaMN for the BEAT dataset. This leads to 12 distinct side-by-side comparison combinations. We randomly select 24 speech segments from the test sets of the ZEGGS and BEAT datasets, respectively, and extract a 10-second clip from each segment starting at an arbitrary position. They are employed in the generation of gestures, yielding 24 video clips for each method. Consequently, there are 288 video pairs in total (24 speech segments $\times$ 12 combinations). Participants are tasked with evaluating 24 video pairs, covering all 24 speech segments. Each of the 12 comparison combinations is presented twice. The orders of both speech samples and the pairing of comparisons are randomized for each subject.

\section{Quantitative Evaluation of LLM-Based Retrieval Model}
\label{sec:details_of_llm_quantitative}

We evaluate the annotation results quantitatively using three methods by two metrics: a) \emph{Accuracy}, which assesses whether the annotated gestures are in the SeG dataset, and b) \emph{Semantic Matching}, which examines whether the annotated gestures at a specific point match the context and semantics of the situation.

\begin{figure}[htbp]
  \centering
  \includegraphics[width=0.8\linewidth]{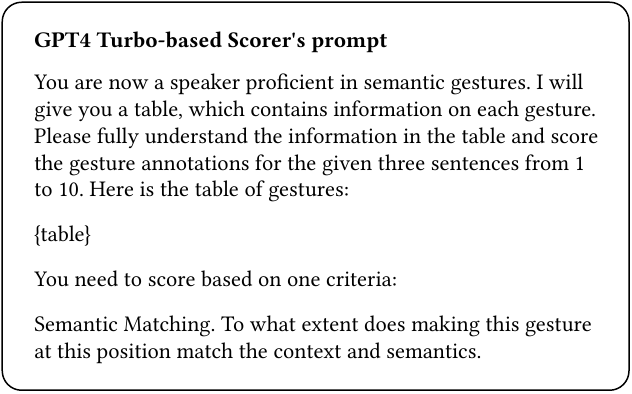}
  \caption{The prompt of GPT4-based Scorer for \emph{Semantic Matching}.}
  \label{fig:gpt4prompt}
\end{figure}

In particular, we randomly select $49$ different speech fragments from the BEAT dataset at first, extracting complete sentences from these fragments' first 20 seconds of transcript. Subsequently, we apply zero-shot, few-shot, and finetuned LLM strategies to perform semantic gesture retrieval on these sentences. For \emph{Accuracy}, we employ word matching to detect whether each annotated semantic gesture appears in the SeG dataset we are given. We calculated this metric as the ratio of correctly annotated semantic gestures to the total number of annotated gestures. For \emph{Semantic Matching}, we utilize the most powerful language model to date, GPT-4 Turbo~\cite{openai2023gpt4}. By employing a prompt-based approach, we make it to play the role of scorer, with each metric rated on a scale from $1$ to $10$. The corresponding prompt is shown in fig \ref{fig:gpt4prompt}. Finally, we calculate the mean values of the three different methods across the two metrics. The average scores are detailed in table \ref{tab:quantitative_evaluation_for_llm}:

\begin{table}[htbp]
    \centering
    \caption{Quantitative evaluation on annotation results. This table reports the mean (± standard deviation) values for each metric by synthesizing on the test data 10 times.}
    \label{tab:quantitative_evaluation_for_llm}
    \begin{tabular}{lccc}
        \toprule
        Metric & zero-shot & few-shot & finetune \\
        \midrule
        Accuracy & $56.7\% \pm 8.1\%$ & $92.8\% \pm 3.8\%$ & $\bm{97.8\% \pm 1.4\%}$\\
        Semantic Matching & $4.35 \pm 0.21$ & $6.00 \pm 0.30$ & $\bm{7.38 \pm 0.29}$ \\
        \bottomrule
    \end{tabular}
\end{table}

As indicated by the qualitative analysis in Section \ref{subsubsec:llm_based_retrieval_model}, the zero-shot strategy performs significantly worse, while the few-shot strategy shows marked improvement in both two metrics, particularly in \emph{Accurancy}. The fine-tuned LLM exhibits a 23.0\% enhancement in \emph{Semantic Matching} beyond the few-shot strategy, reflecting significant advances in semantic comprehension. These results demonstrate the superiority of our fine-tuned LLM.

\end{document}